\documentclass[
superscriptaddress,
preprint,
amsmath,amssymb,
aps,
floatfix,
showkeys,
]{revtex4-2}

\usepackage{booktabs}
\usepackage{color}

\usepackage{graphicx}
\usepackage{dcolumn}
\usepackage{bm}
\usepackage[mathlines]{lineno}
\usepackage{siunitx}

\begin{document}

\title{Orbital selective commensurate modulations of the local density of states in ScV$_6$Sn$_6$ probed by nuclear spins}

\author{R. Guehne}
\email{robin.guehne@cpfs.mpg.de}
\author{J. Noky}
\author{C. Yi}
\author{C. Shekhar}
\affiliation{Max Planck Institute for Chemical Physics of Solids, 01187, Dresden, Germany}
\author{M. G. Vergniory}
\affiliation{Max Planck Institute for Chemical Physics of Solids, 01187, Dresden, Germany}
\affiliation{Donostia International Physics Center, 20018 Donostia - San Sebastian, Spain}
\author{M. Baenitz}
\author{C. Felser}
\affiliation{Max Planck Institute for Chemical Physics of Solids, 01187, Dresden, Germany}

\begin{abstract}
\noindent
The Kagome network is a unique platform in solid state physics that harbors a diversity of special electronic states due to its inherent band structure features comprising Dirac cones, van-Hove singularities, and flat bands. Some Kagome-based non-magnetic metals have recently been found to exhibit favorable properties, including unconventional superconductivity, charge density waves (CDW), switchable chiral transport, and signatures of an anomalous Hall effect (AHE). The Kagome metal ScV$_6$Sn$_6$ is another promising candidate for studying the emergence of an unconventional CDW and accompanying effects. We use $^{51}$V nuclear magnetic resonance (NMR) to study the local properties of the CDW phase in single crystalline ScV$_6$Sn$_6$, aided by density functional theory (DFT). We trace the dynamics of the local magnetic field during the CDW phase transition and determine a loss in the density of states (DOS) by a factor of $\sqrt{2}$, in excellent agreement with DFT. The local charge symmetry of the V surrounding in the CDW phase reflects the commensurate modulation of the charge density with wave vector $q=\left(\frac{1}{3},\frac{1}{3},\frac{1}{3}\right)$. An unusual orientation dependent change in the NMR shift splitting symmetry, however, reveals orbital selective modulations of the local DOS.
\end{abstract}

\keywords{Kagome lattice, charge density wave, nuclear magnetic resonance}

\maketitle

\section*{Introduction}

\noindent
Covalent two-dimensional metals with triangular motifs and confined electronic states are a rich playground in modern solid states physics. Among the non-magnetic materials, systems that are based on Kagome layers attract increasing attention as they naturally feature Dirac cones, van-Hove singularities, and flat bands in their electronic band structure, that may further chemically be tuned to the vicinity of the Fermi level \cite{Wilson2023}. 
Lately, the V based Kagome metal CsV$_3$Sb$_5$ appeared in the spotlight as it undergoes a charge density wave (CDW) phase transition around \SI{94}{K} before it becomes an unconventional superconductor below about \SI{2}{K}, providing the prospect to learn more about the interplay of both phenomena. Evidence for a manipulable chiral transport as well as an anomalous Hall effect (AHE) in the CDW regime further add to the rather diverse list of special electronics of this Kagome metal \cite{Guo2022,Mielke2022}.

Recently, the closely related bilayer Kagome material ScV$_6$Sn$_6$ was reported to undergo a CDW transition at about \SI{92}{K}, including evidence for topologically non-trivial bands and signatures of an anomalous Hall effect, while superconductivity was not observed at ambient pressure \cite{Zhang2022,Arachchige2022,Gu2023,Tuniz2023,Hu2023b,Tan2023,Pokharel2023,Korshunov2023,Kim2023,Cao2023,Shrestha2023,Yi2024,Cheng2024,Lee2024,Subedi2024,Hu2024,Yang2024,Wang2024,Cheng2024b,Zheng2024,Guguchia2023}.
In particular the unconventional nature of CDW phase with a 3-dimensional wavevector $\mathbf{q}_3=\left(\frac{1}{3},\frac{1}{3},\frac{1}{3}\right)$ gives rise to an ongoing debate about the formation mechanism of the charge order \cite{Arachchige2022,Gu2023,Tuniz2023,Hu2023b,Tan2023,Pokharel2023,Korshunov2023,Kim2023,Cao2023,Shrestha2023,Cheng2024,Lee2024,Subedi2024,Hu2024,Yang2024,Wang2024}. Some experiments and theory further suggest a high temperature short range CDW with $\mathbf{q}_2=\left(\frac{1}{3},\frac{1}{3},\frac{1}{2}\right)$ that is unstable at low temperatures \cite{Tan2023,Pokharel2023,Korshunov2023,Cao2023,Wang2024}. 

The complexity of the CDW as well as numerous experimental reports of unusual phenomena led us to employ nuclear magnetic resonance (NMR) to characterize the CDW phase of single crystalline ScV$_6$Sn$_6$ and to search for signatures of an unusual magnetism. Given its high sensitivity for chemical and electronic properties, as well as the atomic resolution, NMR has a long tradition as a method to investigate charge ordered systems, as e.g., in the case of the famous dichalcogenides \cite{Butz1992}. In the context of cuprates, NMR is used to deepen the understanding of the interplay between charge order and superconductivity \cite{Venditti2023,Vinograd2021}. Similarly, NMR of CsV$_3$Sb$_5$ has led to crucial insights into the charge modulation at ambient and elevated pressure \cite{Zheng2022,Luo2022,Song2022,Mu2021,Wilson2023}.

With this paper we present a comprehensive single crystal  $^{51}$V NMR study of the Kagome metal ScV$_6$Sn$_6$ aided by density functional theory (DFT).  We explore the dynamic properties of the local magnetic field during the CDW phase transition between \SI{96}{} and about \SI{80}{K}, and determine a drop in the density of states (DOS) by a factor of $\sqrt{2}$. The CDW phase is characterized as a commensurate, threefold modulation of the local V charge symmetry in accordance with the reported reconstruction of the unit cell with periodicity $\sqrt{3}\times\sqrt{3}\times3$. Our NMR shift data further reveal a peculiar phase shift $\Delta\varphi=\pi/2$ of the sinusoidal modulation of the local magnetic field between the in-plane and the out-of-plane field orientation. This observation can be explained by orbital selective modulations of the local DOS resulting in a distinct wave vector for the V 3$d_{z^2}$ orbitals with $q_{z^2}=\left(\frac{1}{3},\frac{1}{3},\frac{2}{3}\right)$. Finally, we neither find direct evidence of an unusual magnetism as related to time-reversal symmetry breaking and an AHE, nor of an additional, high temperature charge modulation with $\mathbf{q}_2=\left(\frac{1}{3},\frac{1}{3},\frac{1}{2}\right)$.\\

\begin{figure}[t]
\centering
\includegraphics[width=.5\textwidth]{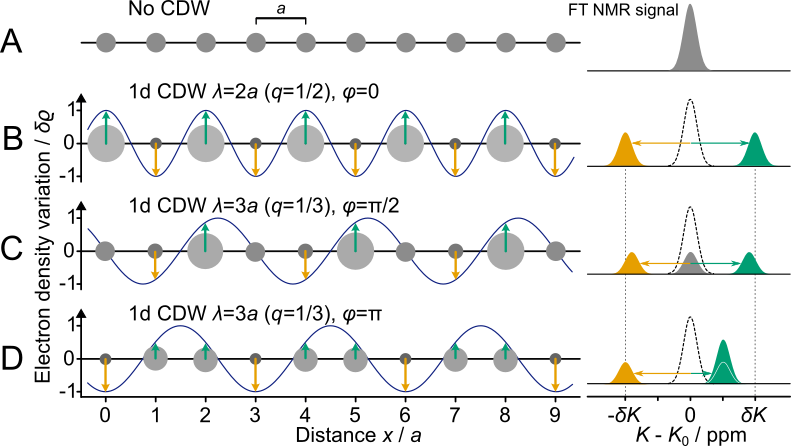}
\caption{\footnotesize{\textbf{One-dimensional commensurate CDWs and NMR}\label{Fig1} (\textsf{A}) A chain of equidistant atoms with evenly distributed electron density (gray circles) yields a single NMR signal at $K_0$ (including intrinsic broadening). (\textsf{B}) For a CDW of amplitude $\delta\rho$, periodicity $q=1/2$, and phase $\varphi=0$, local charge density peaks and valleys alternate between neighboring atoms, yielding a symmetric splitting into two resonance lines at $\pm\delta K\propto\delta\rho$. The signal intensity is proportional to the number of nuclei with the same local environment. (\textsf{C}) A CDW with $q=1/3$ yields 3 equally intense resonance lines. For $\varphi=\pi/2$, two of them are symmetrically shifted to higher and lower frequencies, while the third one remains un-shifted (no extra charge). (\textsf{D}) For the special case of $\varphi=\pi$, two of the 3 lines overlap at $\delta K/2$, while the third one is shifted by $\delta K$ in the opposite direction.
}}
\end{figure}

\noindent
\textbf{Charge density waves and NMR} In the 1960s, Overhauser considered a sinusiodal modulation of the local spin and charge density in metals which is translated into characteristic spectral changes of the corresponding NMR signal through the variation in the NMR shift or quadrupole interaction (as will be discussed in more detail below) \cite{Follstaedt1976}. Consider a one-dimensional (1D) sinusiodal modulation of the local electronic density with wavelength $\lambda=2\pi/q$,
\begin{equation}
\rho=\rho_0+\delta\rho\cos(q\cdot x+\varphi)\ .
\end{equation}
Here, $\rho_0$ denotes the average or un-modulated charge in the CDW regime, $\delta\rho$ the amplitude, $q$ the wave vector, $x$ the real-space position, and $\varphi$ the phase of the CDW. That is, different to most other methods, NMR can also determine the phase $\varphi$ of a charge modulation which can be an important detail in terms of the dynamics of CDWs \cite{Gruner1988}. A variety of commensurate CDW scenarios beside the corresponding NMR signals are illustrated in Fig.~\ref{Fig1}. We assume a periodic lattice of atoms with lattice parameter $a$ and a uniform local electron density $\rho_0$ (panel \textsf{A}). The NMR signal appears at a $K_0$ that corresponds to the uniform local charge. In panels (\textsf{B}) to (\textsf{D}) of Fig.~\ref{Fig1} we show three different CDWs and their NMR resonance patterns. The charge density variation is represented by the size of the circles. The NMR line splits with a maximum of $2\delta K\propto\delta\rho$. Here, NMR shifts to higher and lower frequencies display the gain or loss of electron density for the respective crystal position. In general, the sign of $\delta K$ does not allow a conclusion on whether the corresponding nucleus sits in a valley or on a peak of the CDW, as it depends on the type of hyperfine interaction \cite{Bennett1970}.
For $q=1/2$ (as for $^{51}$V NMR in CsV$_3$Sb$_5$), the resonance line always splits symmetrically, independent of the CDW phase. Contrastingly, for $q=1/3$, as in case of $^{51}$V NMR of ScV$_6$Sn$_6$, the relative position of the three resonances depends on $\varphi$ (cf. panels \textsf{C} and \textsf{D}). This simple model is easily extended to incommensurate CDWs that give rise to broadened NMR spectra with a characteristic double-peak structure at $\pm\delta K$.\cite{Follstaedt1976,Vinograd2021} \\

\begin{figure}[t]
\centering
\includegraphics[width=.5\textwidth]{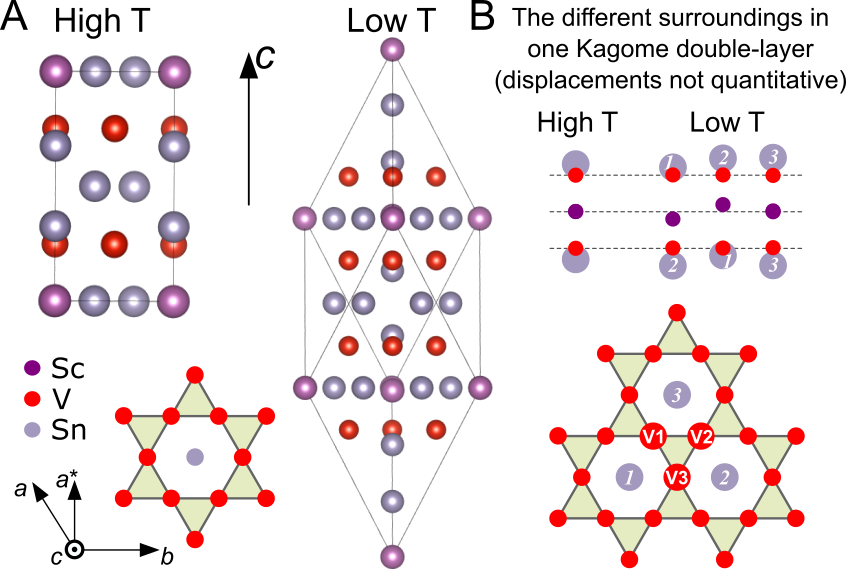}
\caption{\footnotesize{\textbf{The crystal structures of ScV$_6$Sn$_6$}\label{Fig2} (\textsf{A}) On the left, side view of the high temperature unit cell of ScV$_6$Sn$_6$, and below, top view of the V Kagome lattice (red). In the CDW phase, the relative position of the Sn-Sc-Sn chains (gray and purple) is altered along the crystal $c$-axis, yielding 3 different Sc-Sn-Sc units as depicted in panel (\textsf{B}) containing one V Kagome double layer each. This expands the unit cell by a factor of three. The displacement affects the distance of the Kagome-close Sn atoms (\textit{1},\textit{2},\textit{3}), creating 3 distinct V crystal sites (V1, V2, V3).
}}
\end{figure}

\noindent
\textbf{The structural phases of ScV$_6$Sn$_6$} The high temperature phase of ScV$_6$Sn$_6$ has P6/mmm symmetry. The unit cell consists of 13 atoms, where chemically equivalent Vanadium atoms form the Kagome planes, cf. Fig.~\ref{Fig2}~(\textsf{A}). Below about \SI{96}{K}, ScV$_6$Sn$_6$ undergoes a structural phase transition yielding a change in the Sc and Sn positions along the crystal $c$-axis, while the V lattice remains almost unperturbed~\cite{Arachchige2022}. In panel (\textsf{B}) of Fig.~\ref{Fig2}, we show the relevant displacements in terms of Sn-Sc-Sn chains (gray-purple-gray) containing two Kagome layers (red). Different to the symmetric arrangement characteristic for the high $T$ phase, the low $T$ phase features 3 Sn-Sc-Sn chains that vary in length and in relative distance to the V Kagome plane. From the viewpoint of the V Kagome lattice, we identify 3 corresponding Sn positions (\textit{1, 2, 3}) that hover just above (below) the hexagonal V voids, creating 3 non-equivalent V crystal sites (V1, V2, V3), each of them neighbored by two different Sn sites as shown in the lower panel of Fig.~\ref{Fig2} (\textsf{B}).

\begin{figure*}[t]
\centering
\includegraphics[width=\textwidth]{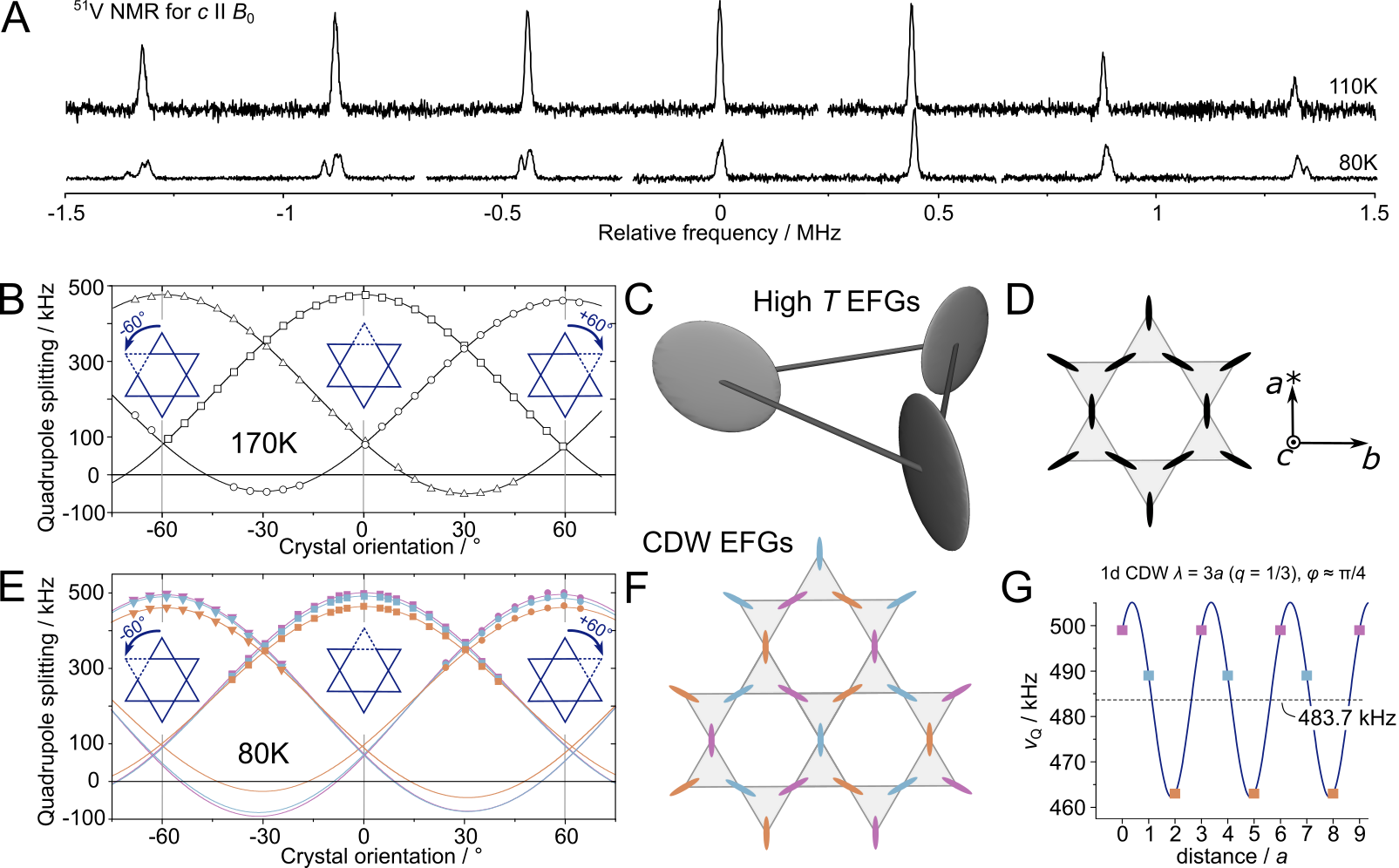}
\caption{\footnotesize{\label{Fig3} \textbf{The local Vanadium charge symmetry for high and low temperatures} (\textsf{A}) The Fourier transform of the $^{51}$V NMR spectra (from selective spin echoes) for \SI{8.73}{T} and $c\parallel B_0$. The spectra were obtained at \SI{110}{} and \SI{80}{K}, confronting the spectral differences between the high and the low temperature phase. The central frequency has been subtracted for the clarity. (\textsf{B}) High temperature orientation dependent quadrupole splittings for rotations about the crystal $c$-axis (in-plane) reveals the six-fold rotation symmetry of the Kagome lattice (cf. Fig.~\ref{S2}). Perspective view (\textsf{C}) and top view  (\textsf{D}) of the equivalent EFGs at V sites in the Kagome lattice. (\textsf{E}) Low temperature orientation dependent quadrupole splittings for in-plane rotations confirms the three sets of quadrupole patterns already observed in panel (\textsf{A}), while the six-fold rotation symmetry of the Kagome lattice is conserved (cf. Fig.~\ref{S4}). (\textsf{F}) Thus, for low temperatures, the three corners of each V triangle become chemically non-equivalent while the EFGs keep the orientation of the high $T$ phase. (\textsf{G}) The commensurate variation of the EFGs along the crystal $a$ or $b$ axis, can be viewed as a 1D sinusoidal modulation with $q=1/3$ and $\varphi\approx\pi/4$.
}}
\end{figure*}

\section*{Results}

\noindent
\textbf{The local Vanadium charge symmetry} The above discussed structural phases feature different local V environments. NMR allows to differentiate them via the electric quadrupole interaction, if such environments have non-cubic symmetry and the nuclear spin satisfies $I>1/2$ ($^{51}I=7/2$). In first order approximation, the quadrupole interaction can be written as
\begin{equation}
\begin{aligned}\label{HQ}
\mathcal{H}_{\mathrm{Q}}&=\frac{3I_z^2-I(I+1)}{4I(2I-1)}eQV_{ZZ}\\
&\times\frac{1}{2}\left(3\cos^2\beta-1+\eta\sin^2\beta\cos\alpha\right)\ ,
\end{aligned}
\end{equation}
where $eQ$ is the nuclear quadrupole moment and $V_{ZZ}$ is the principle value of the electric field gradient (EFG). The EFG is derived from the charge distribution at the nucleus' position, and is thus tightly connected to the above mentioned local environment. The EFG is represented as a traceless ($V_{ZZ}+V_{YY}+V_{XX}=0$), second rank tensor. It is conveniently expressed in terms of its size ($|V_{ZZ}|$) and shape given by the asymmetry parameter $\eta=(V_{XX}-V_{YY})/V_{ZZ}$ ($|V_{ZZ}|\geq|V_{YY}|\geq|V_{XX}|$). The Euler angles $\alpha$ and $\beta$ determine the relative orientation of the EFG's principle axis system with respect to the external magnetic field (typically laboratory $z$-axis), and thus with respect to the sample's crystal lattice.

From \eqref{HQ} it follows that the V nuclear spin system splits into $2I=7$ equidistant resonance lines (note, second order shifts amount to about \SI{1}{kHz} in maximum), 1~central transition (CT) and 3 pairs of satellites, as shown in the top spectrum of Fig.~\ref{Fig3}~(\textsf{A}). The anisotropy of the interaction ($\alpha$ and $\beta$) is transferred to the apparent quadrupolar line splitting $\tilde{\nu}_Q$, i.e., the splitting depends on the sample orientation in the external field, as
\begin{equation}\label{vQ}
\tilde{\nu}_Q=\frac{3eQV_{ZZ}}{4I(2I-1)}\left(3\cos^2\beta-1+\eta\sin^2\beta\cos\alpha\right)\ ,
\end{equation}
with the quadrupole splitting frequency defined as
\begin{equation}
\nu_Q=\frac{3eQV_{ZZ}}{2I(2I-1)}\ .
\end{equation}
By evaluating  $\tilde{\nu}_Q$ for various crystal orientations, size ($|V_{ZZ}|$), shape ($\eta$), and orientation of the EFG can be determined. In Fig.~\ref{Fig3} (\textsf{B}) we show an example of this procedure for the high temperature phase (a detailed account is provided in the supplementary information SI), which corresponds to   $\tilde{\nu}_Q(\beta,\alpha=0)$, i.e., orientation dependent measurements under crystal rotation about the $c$-axis (cf. Fig.~\ref{S2} in SI). We find that $V_{ZZ}$ points along to the crystal $a^{\ast}$-axis with $\nu_Q=\SI{475(2)}{kHz}$ and $\eta=\SI{0.83(1)}{}$. 
The three identical data sets (triangles, squares, and circles) shifted by $\pm\SI{60}{\degree}$ recover the six-fold symmetry of the Kagome lattice (insets) and prove the equivalency of the three corners of each V Kagome triangle in terms of the local Vanadium symmetry.
To visualize the V EFGs in the Kagome lattice, we display them as ellipsoids formed by $|V_{ZZ}|$, $|V_{YY}|$, and $|V_{XX}|$, as shown in panels (\textsf{C}) and (\textsf{D}) of Fig.~\ref{Fig3}.

\begin{small}
\begin{table*}[t]
\caption{\label{Tab1}The electric quadrupole splitting frequency $\nu_Q$, the principle axis of the electric field gradient $V_{ZZ}$ from \eqref{vQ} (for the experiment, only the absolute value can be obtained, the sign cannot be accessed), and the asymmetry parameter $\eta$ as obtained from NMR and $DFT$ at V positions above (High $T$) and below (Low $T$) the CDW phase transition.}
\begin{tabular}{lp{0.3cm}llp{0.3cm}llp{0.3cm}llp{0.3cm}ll}
\toprule
\toprule
& & \multicolumn{2}{c}{High $T$}  & &  \multicolumn{8}{c}{Low $T$}\\
\cmidrule{3-4}\cmidrule{6-13}
& & NMR & \textit{DFT} & & NMR1 & \textit{DFT1} && NMR2 & \textit{DFT2} & & NMR3 & \textit{DFT3}  \\
\midrule
$\nu_Q$ \SI{}{[kHz]}  & & 475(2) & & & 463(2) & & & 489(2) &   & & 499(2) &    \\
$V_{ZZ}$ \SI{}{[10^{20}V/m^2]}  & & 52.9(2) & \textit{-60.812}& & 51.6(2) & \textit{-59.679}& & 54.4(2) & \textit{-61.861} & & 55.6(2) & \textit{-62.617}   \\
$\eta$ & & 0.83(1) & \textit{0.809}& & 0.91(4) & \textit{0.841}  & & 0.81(4) & \textit{0.782}   & & 0.78(3) & \textit{0.787} \\
\bottomrule\bottomrule
\end{tabular}

\end{table*}
\end{small}

Upon cooling, between \SI{96}{} and about \SI{80}{K}, the phase transition takes place and the low temperature structure replaces the high $T$ one. From the NMR point of view (Fig.~\ref{Fig3} \textsf{A} lower spectrum), the changes are most prominently visible in a splitting of the low frequency satellites (find a detailed account on the temperature dependent spectra in Fig.~\ref{ST} in the SI). The single quadrupole pattern observed for high temperature is replaced by 3 sets of quadrupole spectra. In panel (\textsf{E}) of Fig.~\ref{Fig3} we show the results of the same orientation dependent measurements as in panel (\textsf{B}), now performed at \SI{80}{K} (cf. Fig.~\ref{S4} in SI). The change to a threefold quadrupole interaction is consistently found in any possible direction. This proves that the local Vanadium environments have changed, while the relative orientation of the 3 EFGs is same as at high temperatures.
The arrangement of the EFGs is illustrated in panel (\textsf{F}). When viewed along the crystal $a$ or $b$ axis, panel (\textsf{G}), the 3 alternating EFGs create a commensurate 1D sinusoidal variation with $q=1/3$, $\varphi=\pi/4$, and an amplitude of about \SI{20}{kHz} around the average quadrupole splitting of \SI{483.7}{kHz}. Detailed DFT calculations as implemented in VASP~\cite{kresse1996,petrilli1998} confirm the results shown in Fig.~\ref{Fig3}, cf.  Tab.\ref{Tab1}. Note that with the experiment we cannot determine the sign of $V_{ZZ}$ since the quadrupole splitting is symmetric.
\\

\begin{figure}[t]
\centering
\includegraphics[width=.75\textwidth]{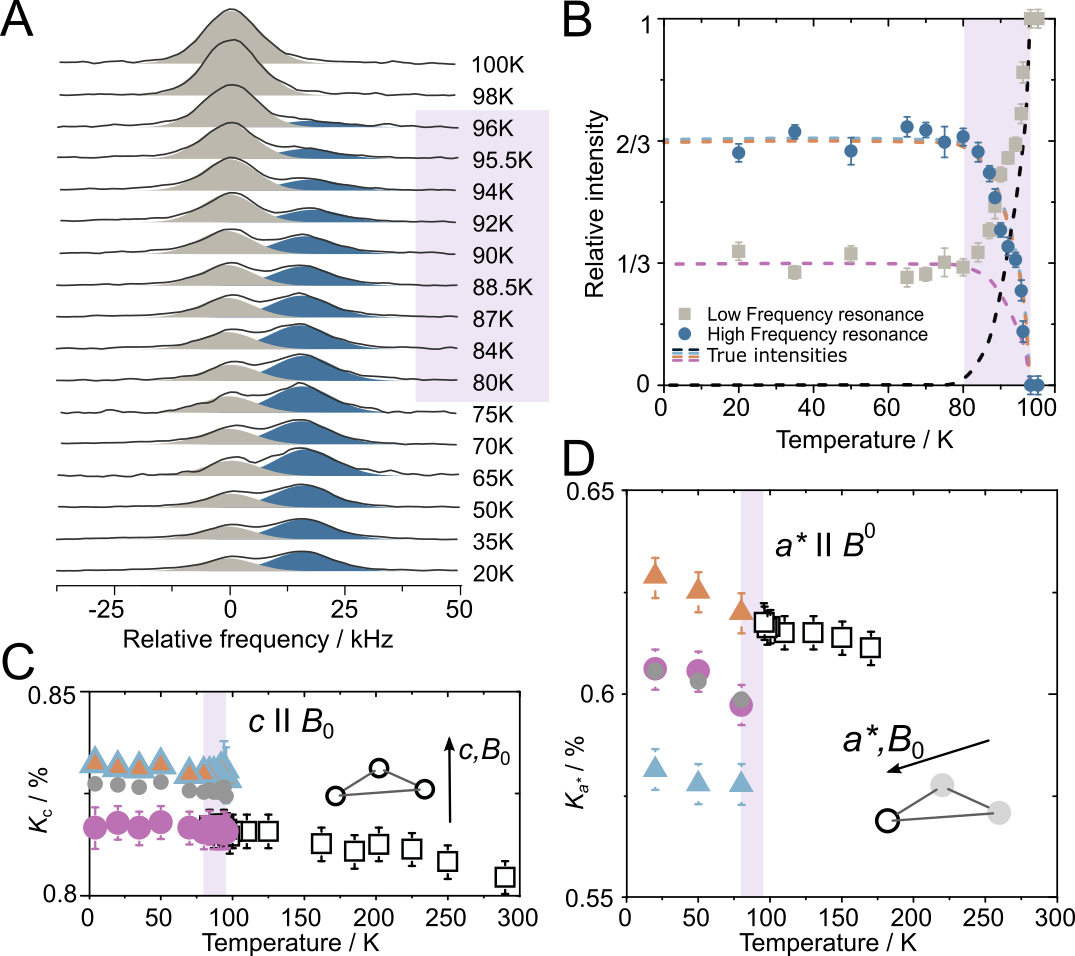}
\caption{\footnotesize{\label{Fig4}\textbf{CDW phase transition and the local magnetic field} (\textsf{A}) The central region of the $c\parallel B_0$ spectrum as function of temperature at \SI{8.73}{T} (Fourier transform of selective spin echoes). Highlighted is the temperature range of the phase transition. Gaussian lines (gray and blue) were used to fit the spectra and to extract the intensity changes. (\textsf{B}) The area of the Gaussians (proportional to signal intensity) from panel (\textsf{A}) as a function of temperature. Dashed lines denote true relative intensities. (\textsf{C}) and (\textsf{D}) The temperature dependence of the NMR shift for $c\parallel B_0$ and $a^{\ast}\parallel B_0$. Note that "$a^{\ast}\parallel B_0$" holds only for 1 of the 3 V nuclei per V triangle as shown in the inset. Gray data points denote the average of the two and three low temperature shift values. 
}}
\end{figure}

\noindent
\textbf{The CDW phase transition} 
In Fig.~\ref{Fig4} (\textsf{A}) we show the center line of the quadrupole spectra from Fig.~\ref{Fig3} (\textsf{A}), for temperatures between \SI{100}{K} and \SI{20}{K}. For the sake of clarity, we have removed the (very small) temperature dependence of the NMR shift. At \SI{96}{K}, first indications of an additional signal (blue) appear about \SI{15}{kHz} ($\sim$\SI{150}{ppm}) next to the initial resonance line (gray). As this shoulder grows in intensity with decreasing temperature, the initial line decreases, until, for about \SI{80}{K} and below, the overall shape of the doublet remains unchanged.

In order to track the volume fractions of the two competing phases during the phase transition, we used Gaussian lines to fit the two resonances in panel (\textsf{A}) of Fig.~\ref{Fig4} and extracted the corresponding signal intensities. The results are shown panel (\textsf{B}). The gray and blue data points denote the respective signal intensities of panel (\textsf{A}) (for an analysis of the total signal intensity, the reader is referred to Fig.~\ref{SQ} in the SI). The individual values are normalized with respect to their sum for each temperature. In the course of cooling, the low temperature resonance line (blue) grows to about 2/3 of the total intensity, leaving about 1/3 (gray) to what appears to be the initial high temperature line. With the colored dashed lines in Fig.~\ref{Fig4} \textsf{(B)} we show the true changes in the relative intensities, concluded from the corresponding changes in the quadrupole splittings. The initial high temperature CT disappears entirely between 96 and about \SI{80}{K} (black dashed line). It is replaced by three resonances (purple, orange, and light blue) with intensity ratio 1:1:1 that represent the 3 CTs of the low temperature quadrupole patterns identified in the former section. Two of the 3 resonances (orange and light blue) form the dark blue peak that appears during the phase transition in panel (\textsf{A}). 

In Fig.~\ref{Fig4}~(\textsf{C}) and (\textsf{D}) we show the temperature dependence of the NMR shift for the out-of-plane orientation of the magnetic field ($c\parallel B_0$) and for the in-plane orientation ($a^{\ast}\parallel B_0$). In the latter case, a reliable determination of the shift values during the phase transition was not possible because of the overlap of high and low temperature resonance lines. For \SI{80}{K} and below, the shift was determined via the satellite transitions (see SI). We assigned the shift values between $c\parallel B_0$ and $a^{\ast}\parallel B_0$ as shown by the colors based on the DFT calculations. The assignment cannot be proven with experiment because the superposition of many resonance lines for orientation dependent measurements prevents a reliable tracing of individual signals. This, however, has no effect on the conclusions of the present investigation.

The V shift in ScV$_6$Sn$_6$ is with more than about \SI{0.6}{\%} and \SI{0.8}{\%} (\SI{6000}{ppm} to \SI{8000}{ppm}) very large, strongly anisotropic, and only very weakly dependent on temperature (2 to \SI{4}{\%} of the total shift).
From detailed orientation dependent measurements (cf. Fig.~\ref{S1} to \ref{S3} in the SI) it can be seen that the high temperature shift tensor is axially symmetric with the main axis pointing towards the center of the V triangle (along the $a^{\ast}$ direction) similar to the EFG.
Although it is very difficult to prove experimentally, it is reasonable to assume that the CDW phase inherits the axial shift anisotropy. Isotropic and axial shift components for the full range of temperatures are depicted in Fig.~\ref{SK} in the SI.

\begin{figure}[t]
\centering
\includegraphics[width=.75\textwidth]{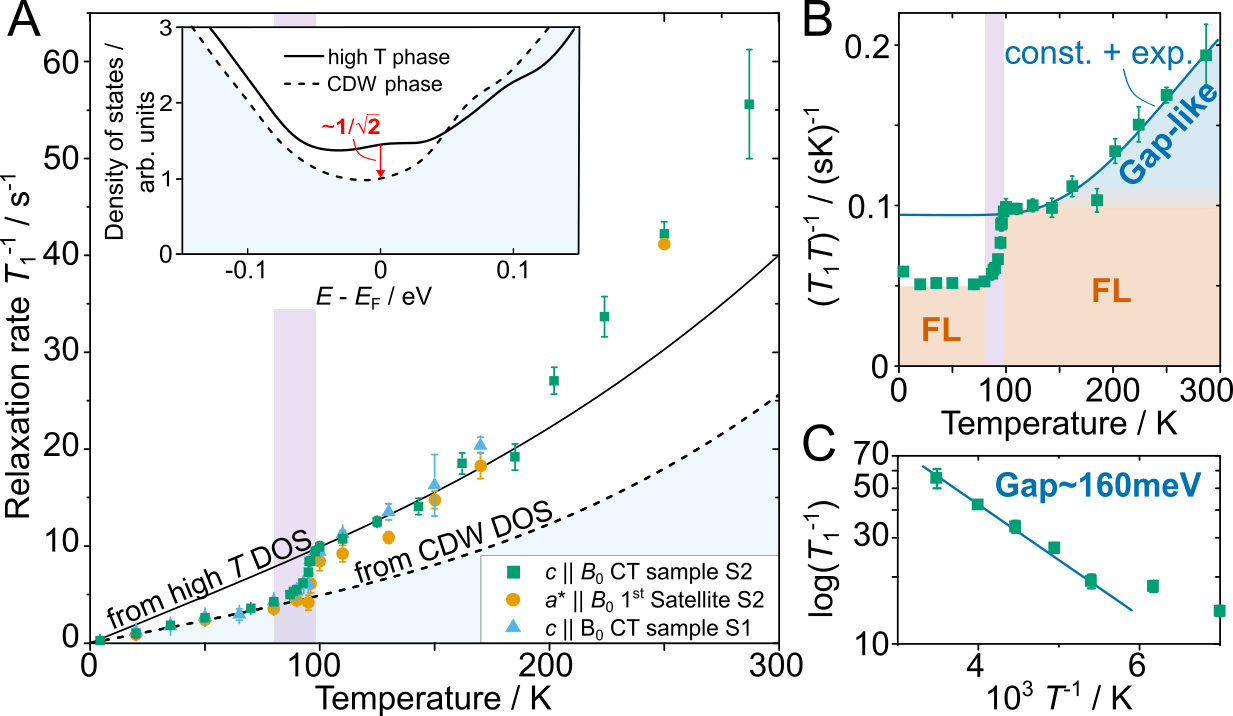}
\caption{\footnotesize{\label{Fig5}\textbf{CDW phase transition and fluctuations of the local magnetic field} (\textsf{A}) The spin-lattice relaxation rate $1/T_1$ as a function of temperature obtained from the CT ($c\parallel B_0$) for sample S1 and S2, and from the first low frequency satellite for $a^{\ast}\parallel B_0$. The inset shows the calculated DOS for V states at high (solid) and low (dashed line) temperatures. At $E_{\mathrm{F}}$, the DOS drops by $\sqrt{2}$ due to the CDW phase transition. (\textsf{B}) $1/T_1T$ as a function of temperature. The blue solid line represents a fit for the high temperature phase data using $\left(const.+\frac{\tilde{T}}{T}\times\exp\left(-\frac{\Delta E/2}{k_BT}\right)\right)$ yielding $\tilde{T}\sim\SI{730}{K}$ and $\Delta E\sim\SI{160}{meV}$. Panel (\textsf{C}) shows the corresponding Arrhenius plot for the high temperature region above \SI{150}{K}. 
}}
\end{figure}

In panel (\textsf{A}) of Fig.~\ref{Fig5} we show the spin-lattice relaxation rate of V nuclei in the two different samples, for two orientations ($c\parallel B_0$ and $a^{\ast}\parallel B_0$), and for a wide range of temperatures. After continuously decreasing between room temperature and about \SI{100}{K}, $1/T_1$ drops abruptly in the region of the phase transition by about a factor of 2. Between \SI{80}{} and \SI{4}{K}, the relaxation rate is proportional to the temperature. Measurements for $a^{\ast}\parallel B_0$ (yellow circles) prove the spin-lattice relaxation to be isotropic (see also Fig.\ref{SMR} in the SI). In the low temperature phase, the data points represent the averaged relaxation of the 3 components. In the inset we present the Vanadium DOS for both phases obtained from DFT. The high temperature DOS are denoted by the solid line, the CDW phase DOS by the dashed line. At the Fermi level, the difference between high and low temperature DOS is $\sqrt{2}$, in agreement with the observed changes in the relaxation (cf. discussion). The temperature dependence of $1/T_1T$, as shown in panel (\textsf{B}), yields a constant value for temperatures below \SI{80}{K} and a pronounced step during the phase transition. For temperatures above $T_{\mathrm {CDW}}$ and $\leq \SI{150}{K}$, $1/T_1T$ appears to be rather constant as well, while it clearly deviates from such a behavior for even higher temperatures. This progressive increase implies a gain in available states as, e.g., in the case of thermal excitation across an energy gap. An activation type of fit (blue line) gives a gap of about \SI{160}{meV}. In panel (\textsf{C}) we show the corresponding Arrhenius plot for the high temperature relaxation rate (above \SI{150}{K}).\\

\section*{Discussion}

\noindent
\textbf{General remarks} The temperature dependent $^{51}$V spectra of ScV$_6$Sn$_6$ as shown in Fig.~\ref{Fig3} (Fig.~\ref{ST} in the SI) and \ref{Fig4} provide consistent evidence of a first order phase transition between \SI{96}{} and about \SI{80}{K} in good agreement with literature \cite{Yi2024}. The low $T$ phase is similarly homogeneous as the high $T$ one, with no additional broadening, neither in the NMR shift nor in the quadrupole interaction, proving any observed changes in the static NMR at low temperatures are perfectly commensurate and sufficiently long range such that domain effects are negligible. Hence, our observations do not support the presence of an additional short-range intermediate charge order \cite{Tan2023,Pokharel2023,Korshunov2023,Cao2023,Wang2024}. The results also do not provide any direct evidence of an unusual magnetism as related to TRS breaking and an anomalous Hall effect. The $c\parallel B_0$-splitting in units of ppm as well as the line shape, i.e., the relative intensity of this double-peak structure, are independent of the applied magnetic field, and thus point at different local spin densities as expected for a CDW (cf. Fig.~\ref{SFD} in SI). Similarly, total signal intensity as well as excitation conditions of the NMR experiments do not indicate any effect from magnetism, as there are no unusual losses or enhancements. That is, any additional effect must be well below the observed linewidth of our resonance lines which is in the order of \SI{100}{ppm}.\\

\noindent
\textbf{Local properties of the CDW phase} The EFG at V sites accessible through the V quadrupole interaction gives a direct evidence of a characteristic charge redistribution due to the CDW phase transition. As shown in the Fig.~\ref{Fig3}, the single high temperature Vanadium EFG is replaced by 3 low temperature EFGs differing in size and shape, but retaining the original orientation in the crystal lattice. The 3 different EFGs mapped on the known low temperature crystal structure correspond to a CDW with wave vector $q=\left(\frac{1}{3},\frac{1}{3},\frac{1}{3}\right)$ which agrees with the reported reconstruction of the unit cell with $\sqrt{3}\times\sqrt{3}\times3$ periodicity. From the splitting frequencies we determine the phase to be $\varphi\approx\pi/4$ for 1D sinusoidal modulations along any crystal direction. The real-space charge distribution that leads to the different EFGs for V1 to V3 cannot be extracted from the experimental results. Plausibly, the EFGs could display the imbalence of Fermi level states in the five V $3d$ orbitals related to the following discussion of the NMR shift. But they may also be affected by charges from bands below the Fermi energy and offsite contributions such as neighbored ions (e.g., the 3 Sn sites in Fig.~\ref{Fig2}) \cite{Guehne2024}.

We further determined the NMR shift for two crystal orientations ($a^{\ast}$ and $c\parallel B_0$) and for a wide range of temperatures as shown in Fig.~\ref{Fig4}~(\textsf{C}) and (\textsf{D}). For the analysis, we may separate the shift into two main components: (1) a very large and highly anisotropic part with a slight temperature dependence, and (2), the characteristic splitting as a consequence of the CDW phase transition. Component (1) is most likely dominated by a very large orbital shift (from \SI{0.6}{\%} to well above \SI{0.8}{\%}) stemming from the magnetic moment associated with the orbital motion of the electrons (van Vleck paramagnetism) that was repeatedly found in metallic V compounds \cite{VanOstenburg1962,Shimizu1963,Drain1964,Clogston1964,Bennett1970,Tsuda1983}. This shift term is typically independent of or weakly dependent on temperature, was reported to be insensitive to the CDW phase transition in 1T-VS$_2$, and may thus explain why the total shift including its temperature dependence seems not to be directly related to the relaxation as expected from a Korringa relation (from $K^2T_1T=const.$ we expect an isotropic shift of less than \SI{0.1}{\%} with a distinct temperature dependence)~\cite{Bennett1970,Tsuda1983}.

\begin{figure*}[t]
\centering
\includegraphics[width=\textwidth]{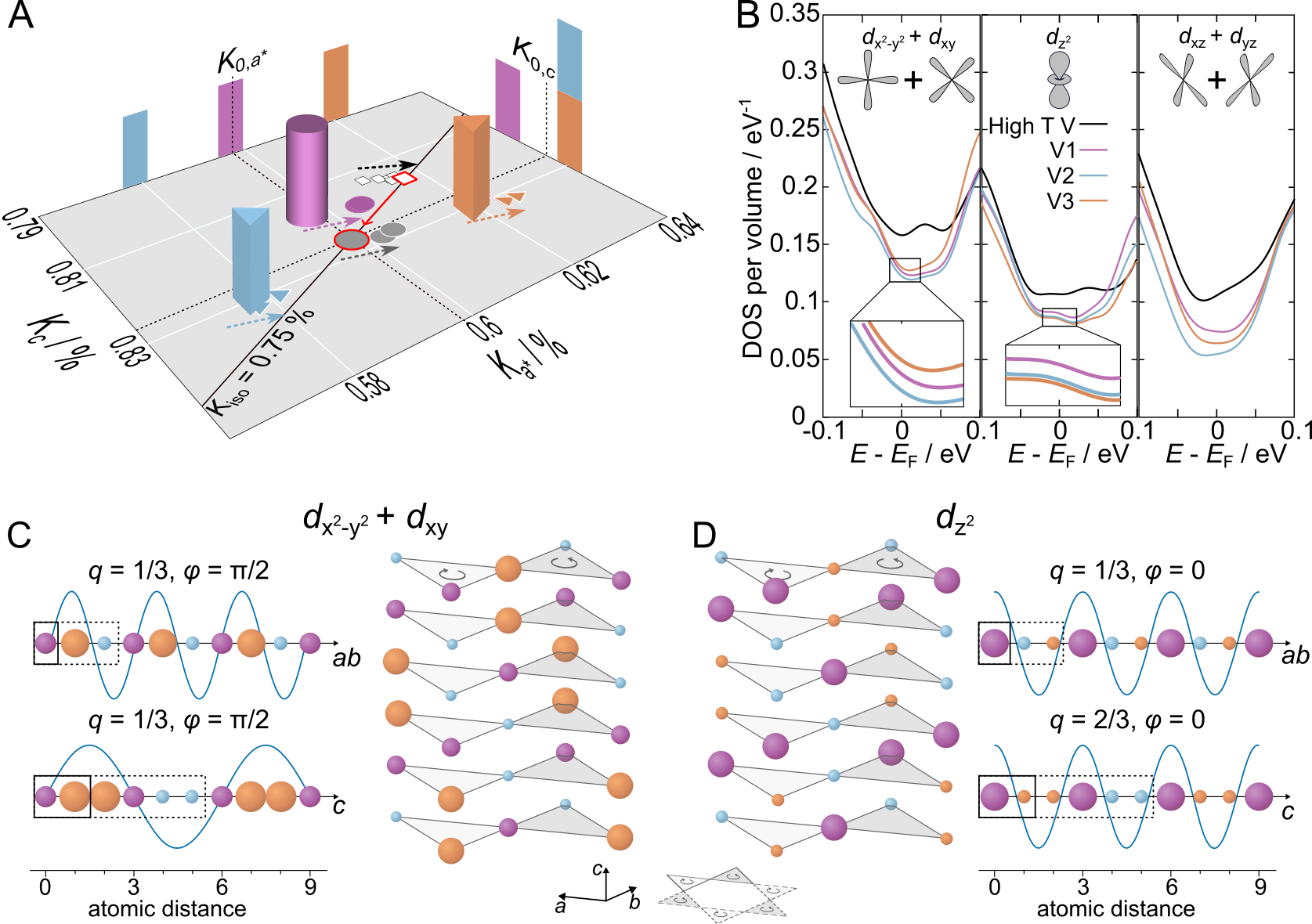}
\caption{\footnotesize{\label{Fig6}\textbf{NMR Shift and the modulation of the local DOS} (\textsf{A}) The shift values from Fig.~\ref{Fig4}~(\textsf{A}) plotted as $K_c$ vs. $K_{a^{\ast}}$ with the temperature as an intrinsic parameter. The red edged black square corresponds to \SI{96}{K}, the red edged gray circle as well as the three 3d objects to \SI{80}{K}. The dashed arrows indicate the direction of $K_i(T)$ for decreasing temperatures. The black diagonal marks a line of constant isotropic NMR shift of \SI{0.75}{\%}. (\textsf{B}) Orbital and site resolved DOS for high (black) and low (colored) temperatures from DFT.  (\textsf{C}) The resulting sinusoidal modulation of the DOS for $d_{x^2-y^2}+d_{xy}$ orbitals ($left$) and the crystal $a$ and $b$ directions (equivalent) as well as for $c$, beside V hourglass building blocks to illustrate the structural stacking of the V Kagome planes in the CDW phase ($right$). Colors denote crystal sites, i.e., V1 to V3, while the size of the balls represent the local DOS. The black solid and dashed boxes denote the unit cell for the high temperature and the CDW structure, respectively. (\textsf{D}) Similar to panel (\textsf{C}) for the $d_{z^2}$ orbitals and the corresponding modulation patterns of the local DOS. Obviously, the wavelength of the DOS modulation for $d_{z^2}$ orbitals is half the CDW unit cell in the crystal $c$-direction, yielding an ${q}_{z^2}=\left(\frac{1}{3},\frac{1}{3},\frac{2}{3}\right)$.}}
\end{figure*}

The second shift component (2) then contains all the changes due to the CDW phase. Obviously, the CDW phase emerges as an in good approximation temperature independent splitting of the single high temperature resonance line into a symmetric triplet of resonances for $a^{\ast}$ and an asymmetric doublet for $c\parallel B_0$. The switching between the two types of splittings appears unusual, but especially the two-fold splitting of \emph{three} components in the $c\parallel B_0$ orientation is surprising, because it is not clear what causes such a symmetry breaking. To illustrate this puzzle, we show in Fig.~\ref{Fig6}~(\textsf{A}) the shift data from Fig.~\ref{Fig4} as a $K_c$ vs. $K_{a^{\ast}}$-plot. The two shift patterns observed for $c$ and $a^{\ast}\parallel B_0$ are sketched as projections on the two perpendicular axes (the 3d objects represent shift data at \SI{80}{K}). They are fingerprints of commensurate 1D CDWs with $q=1/3$ that have a phase difference of $\Delta\varphi=\pi/2$ (cf. Fig.~\ref{Fig1} \textsf{C} and \textsf{D}). The only difference between the two shift splittings is the orientation of the external field and it is rather unlikely that the field directly induces a phase change in the corresponding CDW. This raises the question for what causes the observed phase difference?

The shift anisotropy as a function of temperature brings some light into this problem. Beside the individual temperature dependencies drawn on the plane in Fig.~\ref{Fig6}~(\textsf{A}) (color coding follows Fig.~\ref{Fig4}~(\textsf{C}) and \textsf{D}), the red edged data points, representing $K$ just above (black - \SI{96}{K}) and the averaged $K$ below (gray - \SI{80}{K}) the CDW phase transition, are both located almost exactly on a line of constant isotropic shift of \SI{0.75}{\%}. Under the reasonable assumption of axially symmetric shift tensors for both temperature regimes, the average isotropic shift thus remains unaffected by the phase transition, while the average axial component substantially changes (cf. Fig.~\ref{SK} in the SI). Hence, the observed splittings in $a^{\ast}$ and $c\parallel B_0$ are mainly caused by a splitting in the local electronic spin susceptibility (Pauli paramagnetism) with an anisotropic hyperfine interaction. Contact interaction and core polarization effects that yield isotropic NMR shifts cannot be responsible for the peculiar shift splittings. Spin dipolar interactions, on the other hand side, are highly anisotropic and may offer a way to resolve the conundrum of the phase shift between the two different field orientations \cite{Bennett1970,Boutin2016}. The shift measured for different field orientations could represent the shift contribution from orbitals with different geometry. In particular, the symmetric triplet in the plane may predominately arise from different DOS in the planar orbitals, i.e., $d_{x^2-y^2}$ and $d_{xy}$, while the asymmetric doublet that appears with the field along the crystal $c$-axis displays the DOS of the orbital with a distinct component along that direction, i.e.,~$d_{z^2}$.

To gain independent insight, we examined the DOS of individual orbitals for the 3 different V sites using DFT. The results are presented in Fig.~\ref{Fig6}~(\textsf{B}) resolved for the sum of the $d_{x^2-y^2}+d_{xy}$ orbitals (for symmetry reason both in-plane orbitals are equivalent), for $d_{z^2}$, and for the $d_{xz}+d_{yz}$ orbitals (again, both orbitals are equivalent). The site-selective DOS for the in-plane orbitals at the Fermi level splits symmetrically into 3 different DOS (purple, orange, and light blue), while for the out-of-plane orbital $d_{z^2}$ the splitting is two-fold, with V2 and V3 (orange and light blue) having almost exactly the same DOS. For $d_{xz}+d_{yz}$ the three-fold splitting is again symmetric and visibly larger than for the other two components. Thus, DFT calculations almost perfectly reproduce the characteristic splittings observed with NMR, suggesting that the distinct orbital geometry is dominating the shift for special field orientations. 
Mapping the NMR/DFT data onto the low temperature crystal structure with V1, V2, and V3, we find the periodic, orbital specific modulations of the DOS as shown in panels (\textsf{C}) and (\textsf{D}) of Fig.~\ref{Fig6}. For $d_{x^2-y^2}+d_{xy}$ orbitals, the corresponding wave vector is ${q}_{xy}=\left(\frac{1}{3},\frac{1}{3},\frac{1}{3}\right)$ with a phase $\varphi\approx\pi/2$. The modulation for $d_{xz}+d_{yz}$ orbitals (not shown) has the same wave vector but $\varphi\approx\pi/6$. Due to the twofold splitting in the $d_{z^2}$ orbitals, the corresponding modulation is characterized by ${q}_{z^2}=\left(\frac{1}{3},\frac{1}{3},\frac{2}{3}\right)$ and $\varphi\approx0$. Thus, NMR and DFT imply orbital specific modulations of the DOS that differ in amplitude, phase, and, in case of the V $d_{z^2}$ modulation, even in periodicity of the out-of-plane component. In this scenario, the switching between the in-plane orbitals and the out-of-plane orbitals requires a highly selective hyperfine interaction. This should be subject of a theoretical investigation in future studies. Finally, the chirality of the crystal structure is transferred to the modulations of the DOS where corner-sharing triangular columns (light and dark gray) have opposite handedness. \\

\noindent
\textbf{The CDW phase transition and fluctuations of the local magnetic field} For a metallic system like ScV$_6$Sn$_6$, the spin-relaxation is typically governed by magnetic relaxation, i.e., by fluctuations of the local magnetic field arising from conduction electrons (cf. the corresponding section and Fig.~\ref{SMR} in the SI). The temperature dependence of the relaxation rate $1/T_1$ should then be proportional to the square of the DOS (Korringa relation). We evaluated the following integral,
\begin{equation}
\frac{1}{T_1}\propto A^2_{\mathrm{hf}}\int f(E-\mu_c)[1-f(E-\mu_c)]D(E)^2\mathrm{d}E\ ,
\end{equation}
where $A_{\mathrm{hf}}$ represents the hyperfine interaction that we assume to be temperature independent, $f(E)$ denotes the Fermi function, $\mu_c$ the temperature dependent chemical potential, and $D(E)$ the energy dependence of the DOS for the high and the low temperature phase from the inset of Fig.~\ref{Fig5}~(\textsf{A}). The results are shown by the dashed and the solid black line in Fig.~\ref{Fig5}~(\textsf{A}) after appropriate rescaling. We find a very good agreement between experimental results and DFT. That is, the nuclear spins see the electronic states at the Fermi level and relax via their excitations, while the temperature limits the available states through the Fermi function ($\sim k_{\mathrm{B}}T$). As mentioned in the results section, the relaxation rate values in the CDW phase denote the average of the 3 individual resonance lines, and similarly the low temperature DOS in panel (\textsf{A}) are not resolved for the 3 V sites. It may thus be possible that the relaxation also shows differences among the 3 V sites similar to the shift values. This, however, could not reliably be resolved with experiment.

A different way to look at the relaxation data is presented in Fig.~\ref{Fig5} (\textsf{B}) where the $1/T_1T(T)$ is separated into 3 parts, Fermi liquid behavior below and above the CDW phase transition separated by a change in DOS by a factor of $\sqrt{2}$, and a high temperature dependence above \SI{150}{K} that can very well be approximated by an activation-type of behavior. A single exponential fit yields a gap of the size of about \SI{160}{meV} (cf. Arrhenius plot in panel \textsf{C}) which might be relatable to band structure features, such as van Hove singularities as suggested by others \cite{Mozaffari2023}. In the current case of ScV$_6$Sn$_6$, however, the band structure is too complex to unambiguously relate the spin-lattice relaxation to individual bands and their dispersion. \\

\section*{Conclusions}
\noindent
We investigated the CDW Kagome metal ScV$_6$Sn$_6$ using single crystal $^{51}$V NMR and DFT. The CDW phase transition occurs between \SI{96}{} and about \SI{80}{K} and takes place as a gradual replacement of the high temperature phase by the CDW phase as expected from a first order transition. The phase transition is accompanied by a change in the electronic band structure and the corresponding changes in the total DOS by a factor of $\sqrt{2}$ is very well reproduced with DFT to match the experimentally observed evolution of the local magnetic field fluctuations for decreasing temperatures. The CDW phase features an individual conversion of the three formerly equivalent V environments per V triangle in agreement with DFT and the reported reconstruction of the unit cell with $\sqrt{3}\times\sqrt{3}\times3$ periodicity. In the CDW phase, our findings further comprise an unusual orientation dependent change in the NMR shift splitting from a symmetric triplet of resonance lines for $a^{\ast}\parallel B_0$ to a asymmetric doublet for $c\parallel B_0$, while the latter reflects a symmetry that cannot be found in the crystal structure. When regarded as one-dimensional sinusoidal modulations of the local magnetic field, this observation implies a magnetic field induced phase shift of $\Delta\varphi=\pi/2$. To resolve this discrepancy, we argue on the basis of orbital selective modulations of the local DOS with in-plane wave vector $q=\left(\frac{1}{3},\frac{1}{3}\right)$ but different out-of-plane wave vectors and phases, driven by a field orientation selective hyperfine coupling. The latter calls for a quantitative theoretical treatment of the hyperfine interaction and possible implications for the compound's transport properties. Our work demonstrates that the combination of single crystal NMR and DFT calculations leads to crucial local information about static and dynamic electronic properties in the charge density wave system ScV$_6$Sn$_6$.

\section*{Methods}

\noindent
\textbf{Crystal synthesis} High-quality single crystals of ScV$_6$Sn$_6$ were grown by the flux method \cite{Arachchige2022}. High-purity Sc, V, and Sn elements were loaded in an alumina crucible in a molar ratio of 1~:~10~:~60 and then sealed in a quartz tube under vacuum. The tube was then slowly heated to 1373 K, maintained for 10 h, and cooled down to 973 K over 400 h. Hexagonal shape crystals with silvery surfaces and a typical size of $2\times2\times1\SI{}{mm^3}$ were obtained after centrifugation. The crystal structure was refined from powder x-ray diffraction and the chemical components were examined by using energy-dispersive X-ray spectroscopy The crystal orientations were determined by using the Laue backscattering diffractometer.\\

\noindent
\textbf{NMR experiments} Measurements were carried out on a \textsc{Janis} sweepable \SI{9}{T} magnet and a \textsc{Tecmag Apollo} NMR console. Most experiments were carried out on the hexagonal, plate-like Sample S2 with dimensions $\SI{1.7}{}\times\SI{1.1}{}\times\SI{0.2}{mm^3}$ (cf. Yi et al. 2024 \cite{Yi2024}). A few measurements were done on Sample S1 with $\SI{1.5}{}\times\SI{0.45}{}\times\SI{0.15}{mm^3}$. In both cases, the rf-coil was wound directly around the sample and placed on the single axis goniometer (accuracy $\sim$\SI{1}{\degree}) of a home-built probe. Most experiments were carried out at \SI{8.73}{T} typically using broad-band free induction decay (FID) measurements with pulse lengths of $\SI{0.5}{\mu s}$ in combination with low $Q$-factors (in the order of 10), as well as selective FID or spin-echo ($\pi/2-\tau-\pi$) experiments with $\pi/2$-pulse lengths of 5 or \SI{10}{\mu s} for individual resonance lines ($Q$ between 16 and 32, cf. SI). Selective saturation recovery pulse sequence ($\pi/2-\Delta-$FID/spin echo) was employed to measure the spin-lattice relaxation time $T_1$ of individual transitions.
The shifts were obtained by referencing the $^{51}$V	resonance frequencies versus VOCl$_3$ using the second reference method \cite{Harris2008} and the omnipresent $^{63}$Cu resonance line of the rf-coil.\\

\noindent
\textbf{Numerical calculations} The simulated results were obtained by using \textit{ab initio} calculations in the framework of density-functional theory (DFT), as implemented in the program VASP~\cite{kresse1996}. In this code, augmented plane waves are used as a basis set together with pseudopotentials. To describe the exchange-correlation potential, the generalized-gradient approximation (GGA)~\cite{perdew1996} was used.

The EFGs and self-consistent calculations were carried out on a $18 \times 18\times 9$ ($6\times 6\times 6$) k mesh for the  high (low) temperature phase, respectively. Convergence for total energy and EFGs was carefully checked. For the DOS calculations, a k mesh of $33\times 33\times 33$ ($19\times 19\times 19$) was used for the high (low) temperature phase, respectively.

\section*{Data availability}
\noindent
The data that support the ﬁndings of this study are available from the corresponding
authors upon reasonable request.

\section*{Code availability}
\noindent
The codes that support the ﬁndings of this study are available from the corresponding
authors upon reasonable request.

\section*{Acknowledgment}

\noindent
The authors thank O. Stockert, S. Wirth, J. Sichelschmidt, X. Feng, W. Schnelle, and J. Haase (Leipzig) for helpful discussions. We acknowledge the financial support by the Deutsche Forschungsgemeinschaft (DFG) under SFB1143 (Project No. 247310070), the Würzburg-Dresden Cluster of Excellence on Complexity and Topology in Quantum Matter—ct.qmat (EXC 2147, Project No. 390858490). M.G.V. acknowledges support to the Spanish Ministerio de Ciencia e Innovacion (grant PID2022-142008NB-I00), partial support from European Research Council (ERC) grant agreement no. 101020833, the European Union NextGenerationEU/PRTR-C17.I1, by the IKUR Strategy under the collaboration agreement between Ikerbasque Foundation and DIPC on behalf of the Department of Education of the Basque Government and the Ministry for Digital Transformation and of Civil Service of the Spanish Government through the QUANTUM ENIA project call - Quantum Spain project, and by the European Union through the Recovery, Transformation and Resilience Plan - NextGenerationEU within the framework of the Digital Spain 2026 Agenda. M.G.V. and C.F. acknowledge funding from the Deutsche Forschungsgemeinschaft (DFG, German Research Foundation) for the project FOR 5249 (QUAST).

\section*{Author contributions}
\noindent
R.G performed NMR experiments, analyzed data, and wrote the manuscript. M.B. assisted in the NMR experiments and worked with R.G. on the interpretation of the NMR data. J.N. performed theoretical calculations. J.N. and M.G. assisted in data analysis. C.Y. and S.C. grew and characterized single crystals. C.F supervised the project. All authors commented on the manuscript.

\section*{Competing interest}
\noindent
The authors declare no competing interests.

\section*{Additional information}
\noindent
The Supplementary information is attached at the end of this manuscript.

\bibliography{references}

\newpage
\appendix

\noindent
\large{\textbf{SUPPLEMENTARY INFORMATION:\newline
Orbital selective commensurate modulations of the local density of states in $\textrm{ScV}_6\textrm{Sn}_6$ probed by nuclear spins}}

{\small{
\noindent
R. Guehne,$^1$ J. Noky,$^1$ C. Yi,$^1$ C. Shekhar,$^1$ M. G. Vergniory,$^{1,2}$ M. Baenitz,$^1$ and C. Felser$^1$\\
\textit{$^1$Max Planck Institute for Chemical Physics of Solids, 01187, Dresden, Germany\\
$^2$Donostia International Physics Center, 20018 Donostia - San Sebastian, Spain}}}\\[0.5cm]

\noindent
The following supplement comprises additional experimental data that constitutes essential complementary evidence to the already documented results which did not make it into the main file for reason of space. We will first present temperature dependent spectra for the $c\parallel B_0$ orientation, detailed orientation dependent NMR spectra that allows us to evaluate the electric field gradient (EFG) at the V nuclei for high and low temperatures,  as well as the NMR shift anisotropy. We will next show how the NMR shift is extracted for the $a^{\ast}\parallel B_0$ spectrum where the central transitions are hidden. We will then provide evidence that the total signal intensity can be brought into consistent agreement with changes in temperature, circuit performance, and rf-penetration depth. We will further provide some field dependent measurements, as well as a detailed analysis of the high temperature spin-lattice relaxation to conclude on its magnetic origin.

\section{Temperature dependent $^{51}$V spectra for $c\parallel B_0$}

\begin{figure*}[t]
\centering
\includegraphics[width=.7\textwidth]{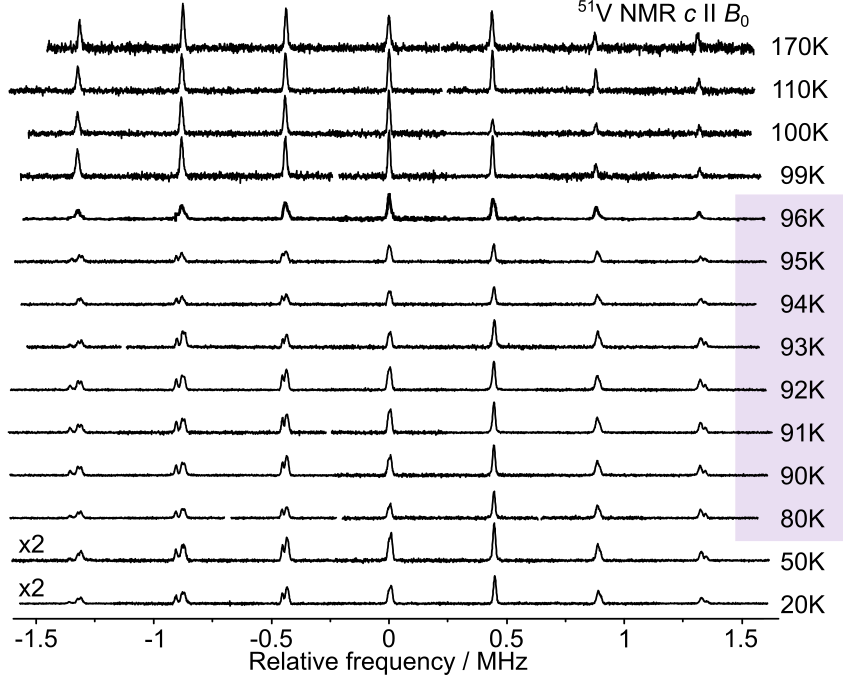}
\caption{\footnotesize{\label{ST} Temperature dependent $^{51}$V NMR spectra for $c\parallel B_0$ and \SI{8.73}{T} (combined Fourier transformed selective spin echoes). The highlighted temperature range marks the phase transition. The peculiar low temperature intensity pattern, especially the pronounced first high frequency satellite, reveal changes in the NMR shift in addition to the obvious differences in the quadrupole splitting frequencies.}}
\end{figure*}

\noindent
In Fig.~\ref{ST} we show a detailed account of the spectral changes for temperatures between \SI{170}{} and \SI{20}{K} and $c\parallel B_0$. The highlighted range of temperatures denotes the CDW phase transition where the high temperature spectrum is progressively replaced by the 3 low temperature spectra.

\section{Evaluation of the EFG tensor at high and low temperatures}
\noindent
In the following Figures \ref{S1} to \ref{S4} the results of detailed orientation dependent measurements are provided. The experiments were carried out for \SI{170}{K} (Figs.~\ref{S1} to \ref{S3}) and  \SI{80}{K} (Fig.~\ref{S4}) at a magnetic field at \SI{8.73}{T} using broad band FIDs ($\SI{0.5}{\mu s}$) and stepwise rotation of the single crystal placed on a single axis goniometer. When rotating the EFG about one of its principle axis ($X,Y$ or $Z$) the apparent quadrupole splitting $\tilde{\nu}_Q$ of the observed spectrum changes according to
\begin{equation}
\tilde{\nu}_Q=\frac{\nu_Q}{2}(3\cos^2\beta-1+\eta\sin^2\beta\cos2\alpha)  
\end{equation}
where $\nu_Q$ denotes the quadrupole splitting frequency defined in the main text and $\alpha$ and $\beta$ are the Euler angles in the PAS of the EFGs.

Note, the various orientation dependent measurements are carried out to consistently determine the orientation of the EFG with respect to the lattice. Size ($V_{ZZ}$) and shape ($\eta$) are obtained from measurements with the single crystal being well adjusted along the $Z$ and $Y$ direction and making use of the EFG being traceless and the definition of the asymmetry parameter.

\begin{figure*}[t]
\centering
\includegraphics[width=\textwidth]{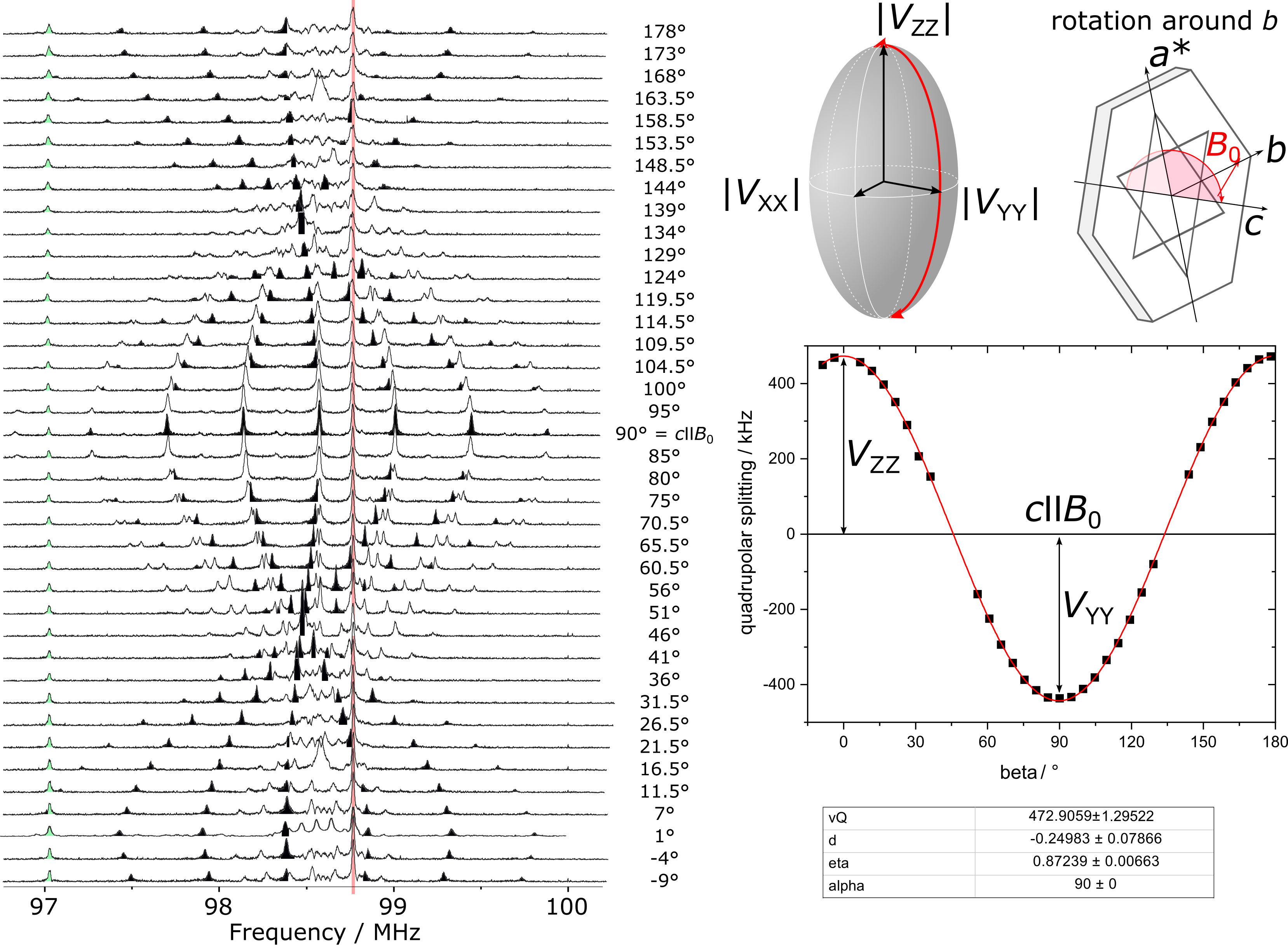}
\caption{\footnotesize{\label{S1} Orientation dependent broad band NMR spectra at \SI{170}{K} and \SI{8.73}{T} for crystal rotations around the crystal $b$-axis which corresponds to a rotation about the $X$-axis of the EFG's PAS (gray ellipsoid). Thus, the rotation connects $V_{ZZ}$ and $V_{YY}$. Note, only the black resonances belong to this well defined rotation, while unmarked resonances represent the other two V nuclei per V triangle that undergo a rotation that leaves the PAS and can thus only be evaluated with much more difficulty. In addition, just above \SI{97}{MHz} and at about \SI{98.75}{MHz} the isotropic $^{27}$Al (light green) and $^{63}$Cu (light red) resonance lines, respectively, can be seen. The signal stem from a small piece of Al metal inside, and from the Cu wire of the rf-coil. The plot on the right hand side shows the quadrupole splitting as function of angle, including the typical fit using the formula above and keeping $\alpha=\SI{90}{\degree}$.}}
\end{figure*}

\begin{figure*}[t]
\centering
\includegraphics[width=\textwidth]{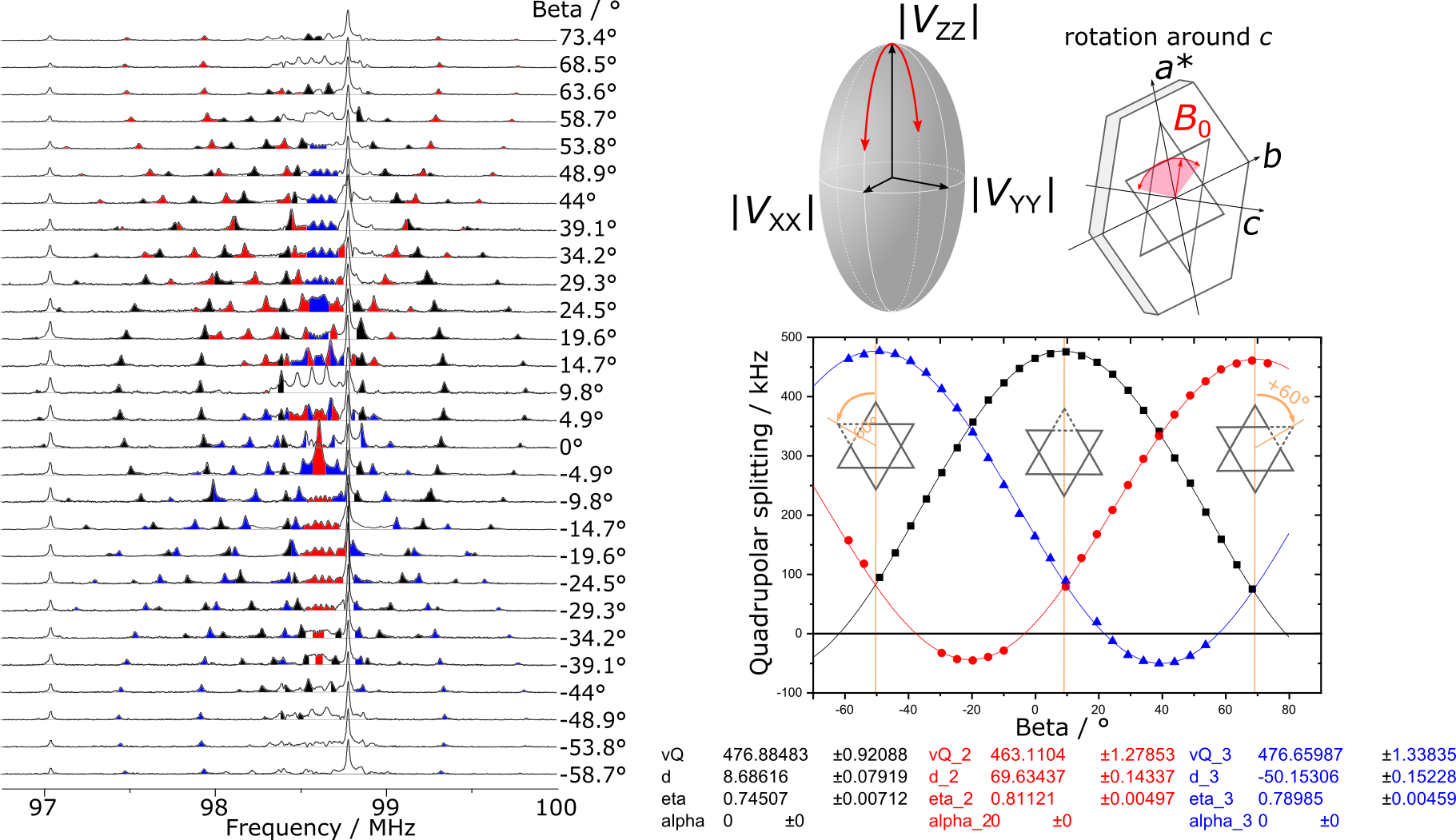}
\caption{\footnotesize{\label{S2} Orientation dependent broad band NMR spectra at \SI{170}{K} and \SI{8.73}{T} for crystal rotations around the crystal $c$-axis which corresponds to a rotation about the $Y$-axis of the EFG's PAS (gray ellipsoid). Thus, the rotation connects $V_{ZZ}$ and $V_{XX}$, the values of which are easily extracted. Note, since the PAS of the EFG's of  each of the three V nuclei per V triangle share the $Z$-axis, all of the resonances (black, blue, and red) can be used to evaluate the EFGs as shown in the plot on the right hand side. Again, the isotropic $^{27}$Al and $^{63}$Cu resonance lines are visible. This time, $\alpha=\SI{0}{\degree}$ and the phase shift $d\approx\pm\SI{60}{\degree}$ determines the relative orientation of the three EFG's and the sixfold symmetry of the Kagome lattice. }}
\end{figure*}

\begin{figure*}[t]
\centering
\includegraphics[width=\textwidth]{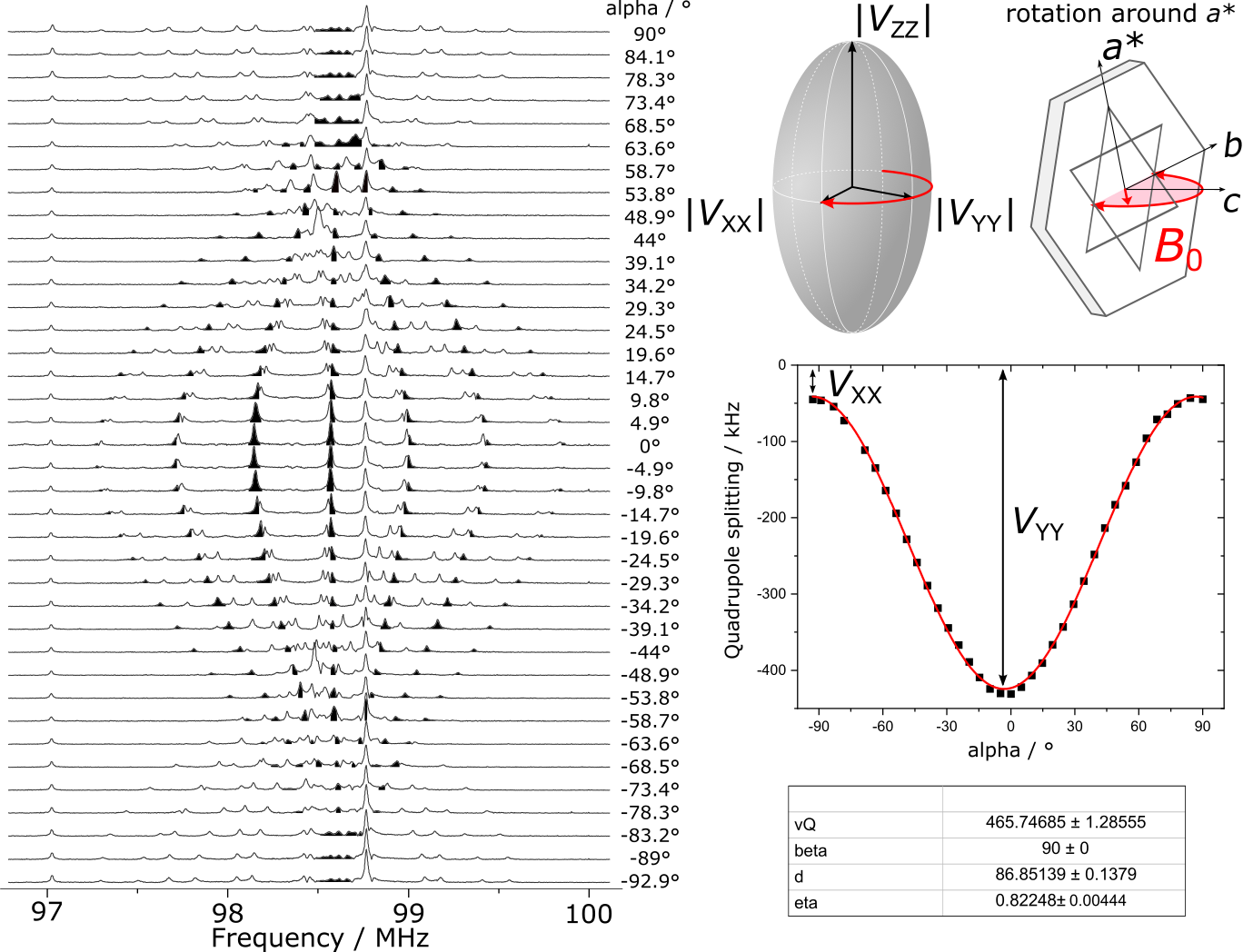}
\caption{\footnotesize{\label{S3}  Orientation dependent broad band NMR spectra at \SI{170}{K} and \SI{8.73}{T} for crystal rotations around the crystal $a^{\ast}$-axis which corresponds to a rotation about the $Z$-axis of the EFG's PAS (gray ellipsoid). Thus, the rotation connects $V_{YY}$ and $V_{XX}$.  Note, only the black resonances belong to this well defined rotation, while unmarked resonances belong to the other two V nuclei per V triangle that undergo a rotation that leaves their PAS and can thus only be evaluated with much more difficulty. The plot on the right hand side shows the quadrupole splitting as function of angle, including the typical fit using the formula above while keeping $\beta=\SI{90}{\degree}$.}}
\end{figure*}

\begin{figure*}[t]
\centering
\includegraphics[width=\textwidth]{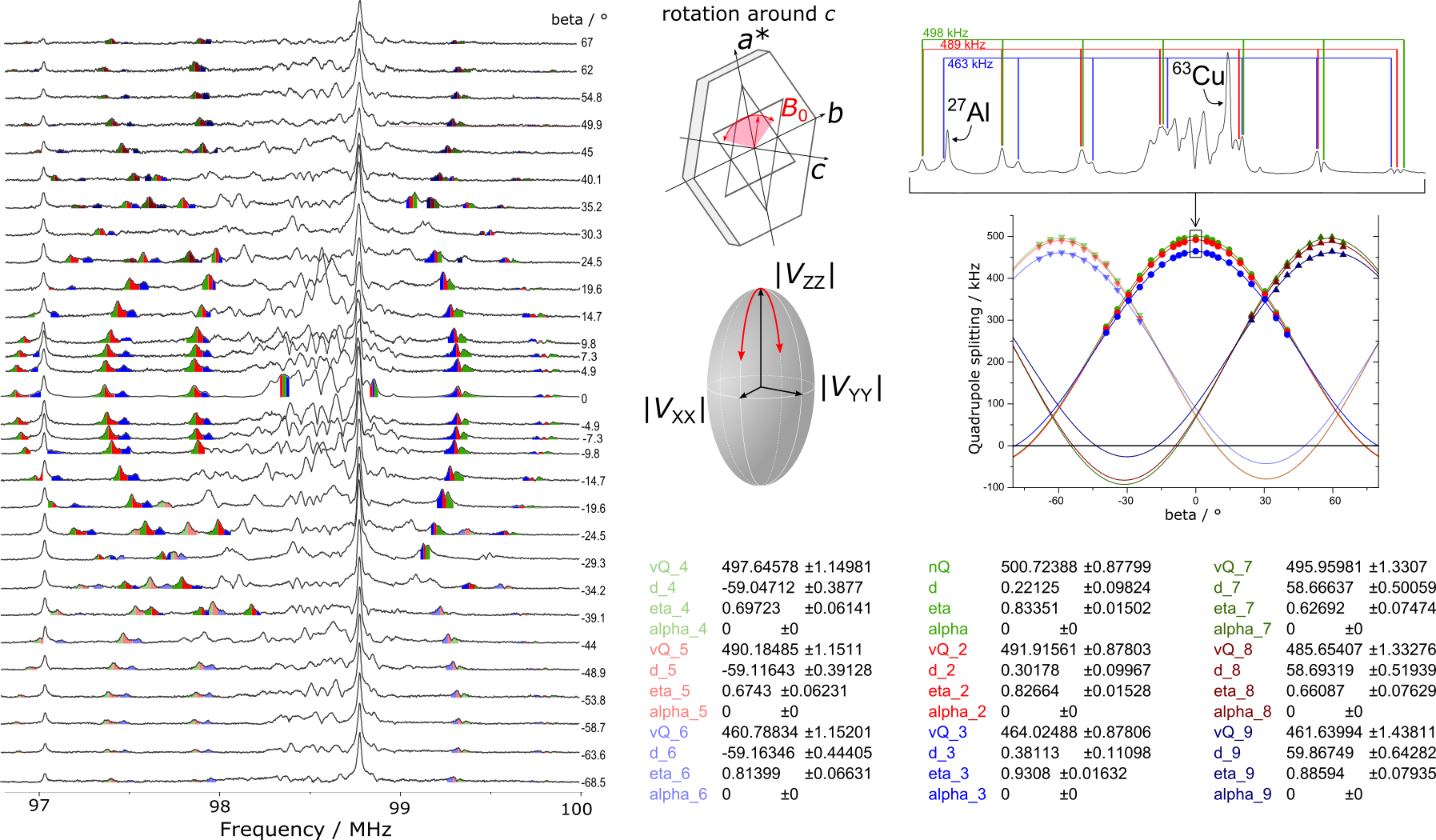}
\caption{\footnotesize{\label{S4} Orientation dependent broad band NMR spectra at \SI{80}{K} and \SI{8.73}{T} for crystal rotations around the crystal $c$-axis similar to the one shown in Fig.~\ref{S2}.  In the CDW phase, the NMR spectrum visible splits into 3 individual spectra for each of the 3 V per V triangle (green, red, and blue $-$ in light, regular, and dark color) while the sixfold symmetry of the Kagome lattice is retained. The plot on the right documents the orientation dependence and sixfold symmetry of the underlaying lattice.}}
\end{figure*}

\section{Determination of the NMR shift for the in-plane orientation}
\begin{figure*}[t]
\centering
\includegraphics[width=.6\textwidth]{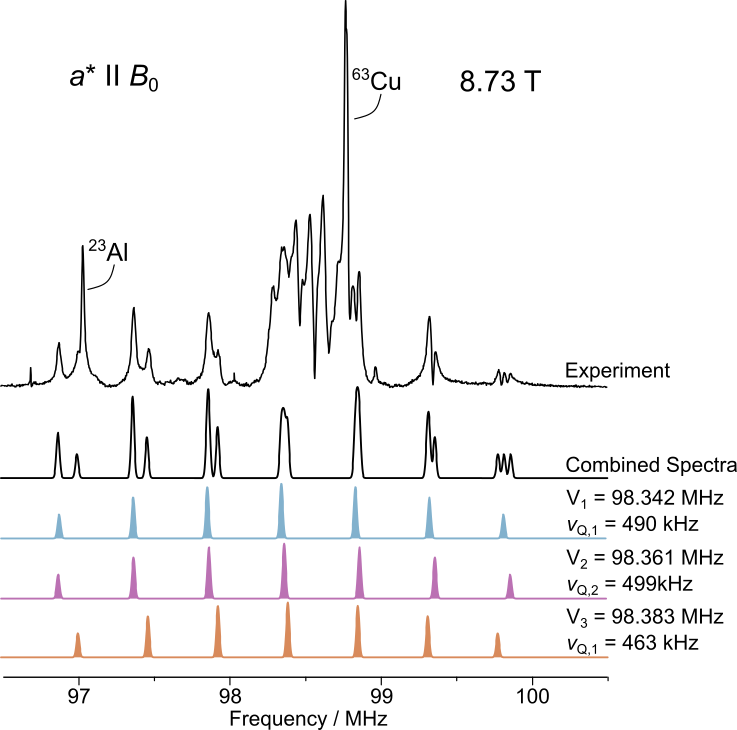}
\caption{\footnotesize{\label{SSpec} Fourier transform of a broad band NMR signal for \SI{80}{K} and \SI{8.73}{T} with the single crystal aligned as $a^{\ast}\parallel B_0$. The four spectra below represent the three individual quadrupole patterns (colored) and their sum (black).}}
\end{figure*}

\noindent
For $c\parallel B_0$ the central transitions and thus the NMR shift are directly accessible as can be seen in Fig.~\ref{Fig4} (\textsf{A}). For the $a^{\ast}\parallel B_0$ orientation, the CTs (for high and low temperatures) are hidden behind the broad quadrupole spectrum of the 2 other V nuclei in each V triangle. In order to obtain the NMR shift, therefore, the satellites can be used, and the first order (equidistant) quadrupole pattern. The procedure is shown in Fig.~\ref{SSpec}. The relevant low temperature spectrum consists of 3 equally intense quadrupole patterns differing in shift and quarupole splitting (purple, orange, and light blue).

\section{NMR shift anisotropy}
\begin{figure*}[t]
\centering
\includegraphics[width=.9\textwidth]{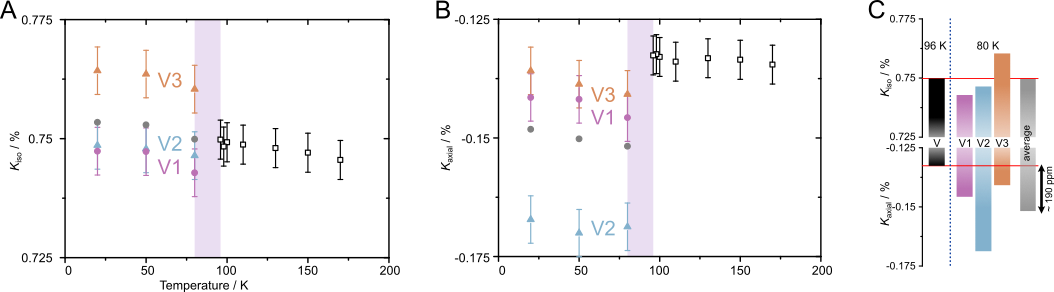}
\caption{\footnotesize{\label{SK} The isotropic and axial shift component as functions of temperature. Note that for low temperatures an axial symmetry cannot be verified with the experiment. The values during the phase transition are missing because for $a^{\ast}\parallel B_0$ it is not possible to identify individual resonances due to the overlap of resonances.}}
\end{figure*}

\noindent
For the high temperatures, the NMR shift was found to be axially symmetric, as there is essentially no change in the resonance frequency of the CT in Fig.~\ref{S3}. It is a reasonable assumption that the CDW phase inherits this symmetry. The corresponding isotropic ($K_\mathrm{iso}=(2K_c+K_{a^{\ast}})/3$) and axial ($K_\mathrm{axial}=2(K_c-K_{a^{\ast}})/3$) shift components are shown in Fig.~\ref{SK}. The colors follow Fig.~\ref{Fig4}~(\textsf{C}) and (\textsf{D}) and are identified with the high single temperature resonance V and three chemically non-equivalent low temperature V sites V1, V2, and V3. In the CDW phase, the isotropic shifts assume 3 different values, from about \SI{0.743}{\%} for V1 to about \SI{0.76}{\%} for V3. Surprisingly, their mean value (gray column in panel \textsf{C}) agrees almost exactly with $K_\mathrm{iso}$ for the high temperature resonance just above the onset of CDW phase transition at \SI{96}{K} (black column). Contrastingly, the anisotropy of all 3 low temperature V sites increases, yielding values between \SI{-0.14}{\%} (V3) to almost \SI{-0.17}{\%} (V2). The mean axial shift component (gray) deviates clearly from the high temperature $K_\mathrm{axial}$ at \SI{96}{K}, with a difference in the order of \SI{190}{ppm} ($\SI{0.019}{\%}$). Thus, one way to look at the CDW transition is through the shift anisotropy, with a net change in the axial part of the shift ($\Delta K_\mathrm{axial}$), while the average isotropic shift remains unaltered in comparison to the high temperature phase.

\section{The circuit's quality factor, rf-penetration depth, and the total signal intensity}

\begin{figure*}[t]
\centering
\includegraphics[width=\textwidth]{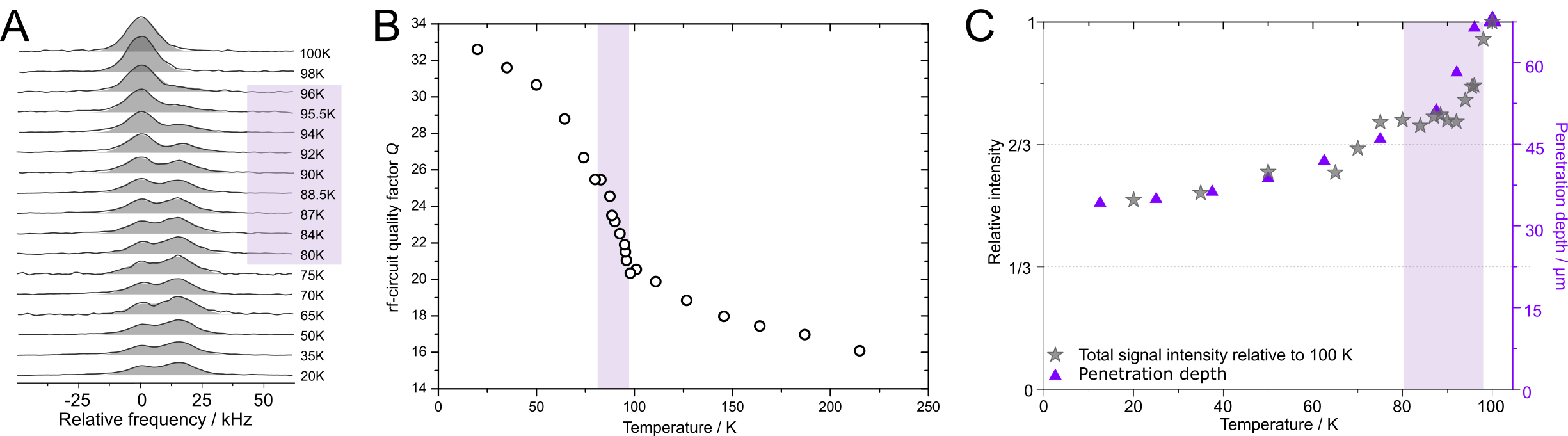}
\caption{\footnotesize{\label{SQ} (\textsf{A}) The CT for $c\parallel B_0$ as a function of temperature reproduced from Fig.~\ref{Fig4}. The gray areas correspond to the total signal intensity. (\textsf{B}) The rf-circuit's quality factor $Q$ as a function of temperature. Clearly, a bend in the dependence is visible at the onset of the CDW phase transition, most potentially due to the corresponding change in the sample resistivity and the thus changing rf-penetration depth that affects the inductance of the rf-coil. (\textsf{C} The CT signal intensity from panel (\textsf{A}) after correction for the temperature and $Q$ fits matches almost perfectly with the $T$ dependence of the rf-penetration depth (rescaled to match the \SI{100}{K} value of the intensity) implying the loss in signal intensity is due to a loss in accessible nuclei from a reduced penetration depth.}}
\end{figure*}

\noindent
To evaluate the total signal intensity  we investigate the central transition of Fig.~\ref{Fig4} (\textsf{A}). This allows us to eliminate some uncertainties that are connected with the excitation conditions in terms of selective excitations, power levels, band widths etc. The CTs are  nevertheless representative for the total signal intensity. We reproduce the spectra in Fig.~\ref{SQ} (\textsf{A}). These spectra are corrected for signal averaging as well as for the temperature. An NMR signal is further proportional to $\sqrt{Q}$, where $Q$ is the quality factor of the rf-circuit. We plot $Q$ as a function of temperature in panel (\textsf{B}). $Q$ changes with temperature since it is connected to the resistivity of the sample. We thus corrected the signal intensity given by the gray area under the curves in panel (\textsf{A}) by the changes in $Q$, and plot it in panel (\textsf{C}) as gray stars. Finally, since the resistivity of the sample changes significantly as a function of temperature \cite{Yi2024}, we estimated the corresponding penetration depth of the rf-field at $f=\SI{100}{MHz}$ due to the skin effect, using $\delta(T)=\sqrt{2\rho(T)/(2\pi f\mu_0)}$, where $\rho$ denotes the sample resistivity, $f$ the frequency, and $\mu_0$ the vacuum magnetic permeability. The results are plotted as purple triangles in panel (\textsf{C}). Consistently, as the resistivity decreases with decreasing temperature, the skin depth decreases as well, thus, reducing the number of nuclear accessible with NMR. Obviously, the total signal intensity's dependence on temperature follows very well the changes of the skin depth expected for a thick sample (\SI{200}{\mu m}). This analysis confirms that we have no unusual intensity loss in the present system.

\begin{figure*}[h!]
\centering
\includegraphics[width=.4\textwidth]{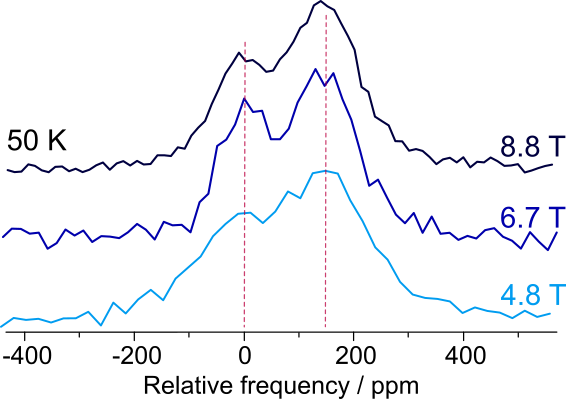}
\caption{\footnotesize{\label{SFD} Fourier transform of spin-echo measurements of the central double-peak for $c\parallel B_0$ and \SI{50}{K} for external fields of 4.8, \SI{6.7}{}, and \SI{8.8}{T} in units of ppm. The red dashed lines denote the approximate center frequencies of the two peaks. A constant splitting in units of ppm proves the origin of the splitting is due to different NMR shifts (paramagnetic spin susceptibility) rather then an unusual magnetism.}}
\end{figure*}
\noindent

\section{Magnetic field dependent measurements}
\noindent
The double-peak central transition structure from Fig.~\ref{Fig4} (\textsf{A}) was measured for two more magnetic fields. The spectra are shown in Fig.~\ref{SFD} in units of ppm. Evidently, the two peaks do not change in relative position (\SI{150}{ppm}), nor in relative intensity. This is a strong evidence that the origin of this peak system is related to different Knight shift values, i.e., different DOS as expected from a CDW. Second order quadrupole effects ($\propto1/B_0$) obviously do not affect the spectrum, while, on the other hand, unusual magnetism as potentially related to orbital currents, are not evident.

\section{Magnetic relaxation and determination of $T_1$}

\noindent
In a quadrupolar split system, it can be difficult to determine $T_1$, as it is defined to be the relaxation time of the unsplit spin system, i.e., as if no quadrupole interaction is present. For $^{51}$V, the quadrupole spectrum covers almost \SI{3}{MHz} and thus, the relaxation of individual transitions is the only way to measure $T_1$ for fairly well defined conditions. The relaxation of individual transition, however, depends on the transition and the relaxation mechanism, i.e., whether its driven by magnetic fluctuations due to free carriers or lattice vibrations via the quadrupole interaction.

In order to check the relaxation mechanism in the current system, we assume the relaxation to be of magnetic origin, and apply the corresponding recovery equations to extract $T_1$. We used selective saturation recovery pulse sequences and analyzed the individual recovery curves with:
\begin{equation}\label{SR}
M(t)=M_0\left[1-f\Biggl\{\sum_{i=1}^7 a_i\cdot\exp\left(-\frac{\lambda_it}{T_1}\right)\Biggr\}\right]\ ,
\end{equation}
where $M_0$ is the equilibrium signal intensity, and $f$ the inversion factor ($f=1$ saturation, $f=2$ inversion). The coefficients $a_i$ vary for each transition, while the exponents $\lambda_i$ are shared. We summarize the various values in Tab.~\ref{TabSR}.
\begin{table}
\centering
\setlength\tabcolsep{8pt}
\caption{\label{TabSR}Coefficients $a_i$ and exponents $\lambda_i$ for magnetic relaxation and recovery measurements of selectively saturated or inverted transitions of a spin $7/2$ system.}
\begin{tabular}{ccccc|r}
\toprule\toprule
$a_i$ & CT & 1st Sat & 2nd Sat & 3rd Sat & $\lambda_1$\\
\midrule
$i=1$ & 1/84 & 1/84 & 1/84 & 1/84 & 1\\
2 &0 & 1/84 & 1/21 & 3/21 & 3\\
3 & 3/44 & 1/33 & 1/132 & 3/11 & 6\\
4 & 0 & 9/77 & 25/308 & 25/77 & 10\\
5 & 75/364 & 1/1092 & 100/273 & 75/364 & 15\\
6 & 0 & 49/132 & 49/132 & 3/44 & 21\\
7 & 1225/1716 & 392/858 & 98/858 & 8/858 & 28\\
\bottomrule\bottomrule
\end{tabular}
\end{table}

\begin{figure*}[t]
\centering
\includegraphics[width=.4\textwidth]{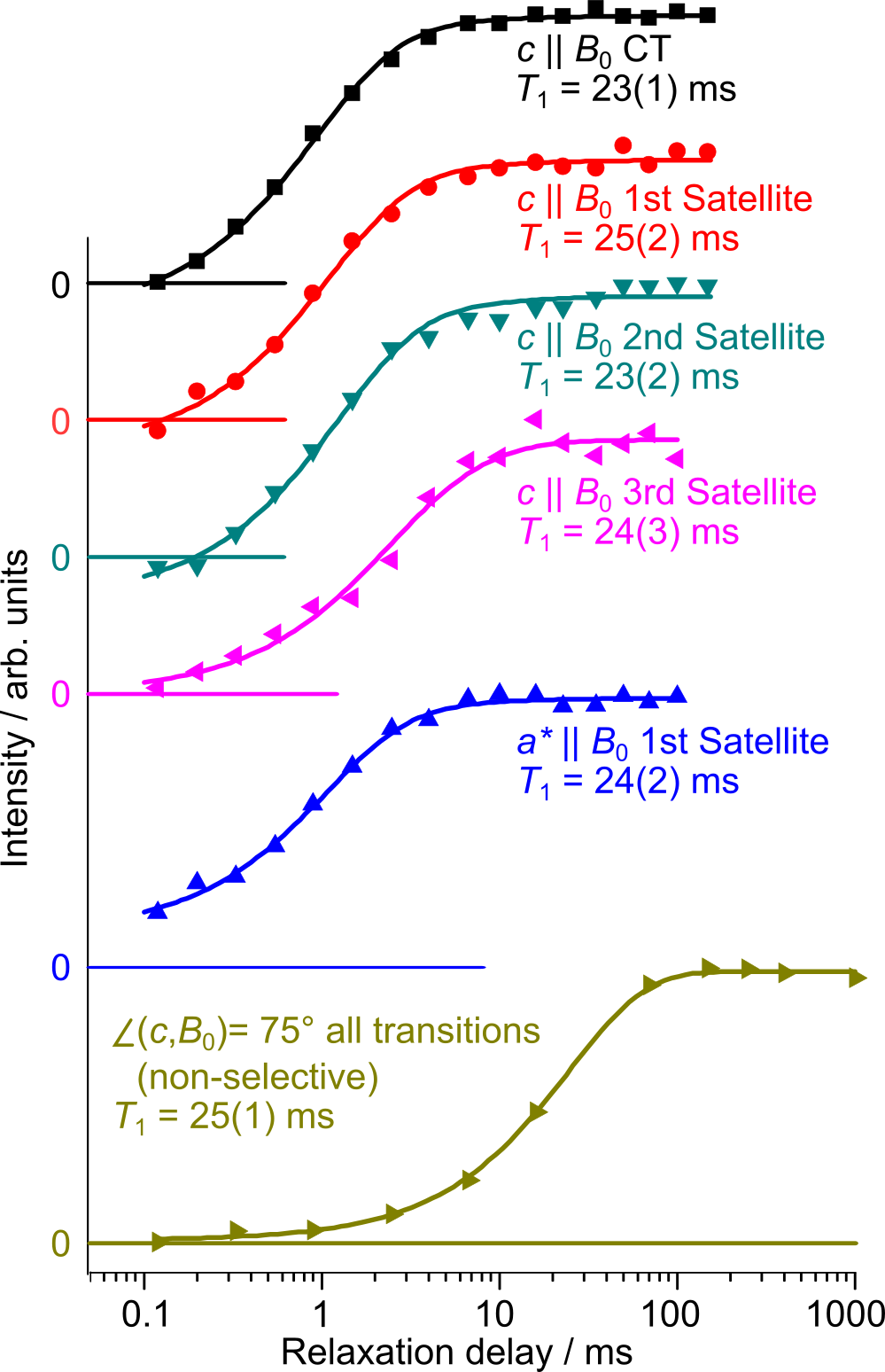}
\caption{\footnotesize{\label{SMR} Evaluation of the spin-lattice relaxation at \SI{250}{K} and \SI{8.73}{T}. The plot shows the results from individual saturation recovery measurements carried out for the CT (black), and the first (red), second (petrol), and third (pink) satellite of the V spectrum for $c\parallel B_0$, as well as for the first satellite (blue) for $a^{\ast}\parallel B_0$, and for the non-selective spectrum at \SI{75}{\degree} (yellow). The solid curves denote the best fit using equations \eqref{SR}, \eqref{SExp}, and Tab.~\ref{TabSR}.}}
\end{figure*}

The results are shown in Fig.~\ref{SMR}. Black denotes the recovery for the CT, red for the first, petrol for the second, and pink for the third  lower satellite. The results provide strong evidence for the assumption, because each of the 4 transition selective recoveries, i.e, CT and the three satellites, yield the same $T_1$ of \SI{24(1)}{ms} at \SI{250}{K} and $c\parallel B_0$. We further repeated the measurement for the 1st satellite at the same temperature for $a^{\ast}\parallel B_0$ (blue data), which yields the same result, and hence, the relaxation is isotropic.

Finally, there are a few crystal orientations where one of the 3 V quadrupole splittings disappears, and therefore, the quadrupole interaction is eliminated. We used $\angle(c,B_0)\approx\SI{75}{\degree}$ (cf. Fig.~\ref{S1}) to measure the pure $T_1$ of the un-split spin system (Fig.~\ref{SMR} yellow data) and applied the simple exponential relationship 
\begin{equation}\label{SExp}
M(t)=M_0\left[1-f\cdot\exp\left(-\frac{t}{T_1}\right)\right]
\end{equation}
which, evidently, gives the same $T_1$ as for selective excitation in the other orientations. We have thus confirmed the relaxation to be driven by magnetic fluctuations from free electrons which is the prerequisite for our treatment of $1/T_1(T)$ using the DOS.

\end{document}